\documentclass[a4paper,fleqn,usenatbib]{mnras}
\usepackage[T1]{fontenc}
\usepackage{ae,aecompl}
\usepackage{csquotes}
\usepackage{amsmath}	
\usepackage{amssymb}	
\usepackage{epsfig}

{\newif\ifnotend
\notendtrue
\def\veclist{ABCDEFGHIJKLMNOPQRSTUVWXYZabcdefghijklmnopqrstuvwxyz.}
\def\top#1#2.{#1}
\def\tail#1#2.{#2.}
\loop\expandafter\xdef\csname v\expandafter\top\veclist\endcsname%
{{\noexpand\bf\expandafter\top\veclist}}
\edef\veclist{\expandafter\tail\veclist}
\if\veclist.\notendfalse\fi\ifnotend\repeat}

\newcommand {\Myr}{\,{\rm Myr}}

\newcommand {\Gyr}{\,{\rm Gyr}}
\newcommand {\K}{\,{\rm K}}

\newcommand {\kpc}{\,{\rm kpc}}

\newcommand {\SF}{{\rm SF}}

\newcommand {\feh}{\hbox{[Fe/H]}}
\newcommand {\mgh}{\hbox{[Mg/H]}}
\newcommand {\eufe}{\hbox{[Eu/Fe]}}
\newcommand {\eualpha}{\hbox{[Eu/$\alpha$]}}

\newcommand {\mgfe}{\hbox{[Mg/Fe]}}
\newcommand {\eumg}{\hbox{[Eu/Mg]}}
\newcommand {\eusi}{\hbox{[Eu/Si]}}
\newcommand {\sife}{\hbox{[Si/Fe]}}
\newcommand {\simg}{\hbox{[Si/Mg]}}

\newcommand {\rfe}{\hbox{[r/Fe]}}
\newcommand {\eu}{{\rm Eu}}
\newcommand {\mg}{{\rm Mg}}
\newcommand {\si}{{\rm Si}}
\newcommand {\fe}{{\rm Fe}}
\newcommand {\afe}{\hbox{[$\alpha$/Fe]}}
\newcommand {\gdfe}{\hbox{[Gd/Fe]}}
\newcommand {\dyfe}{\hbox{[Dy/Fe]}}
\newcommand {\dex}{\,{\rm dex}}

\newcommand {\Msun}{\,{\rm M}_\odot}
\newcommand {\Mch}{{\rm M_{ch}}}

\newcommand {\fcccSN}{\rm f_{\rm c,ccSN}}

\newcommand {\fcNM}{\rm f_{\rm c,NM}}
\newcommand {\ccSN}{{\rm ccSN}}
\newcommand {\fcSNIa}{{\rm f_{\rm c,SNIa}}}
\newcommand {\snIa}{{\rm SNIa}}
\newcommand {\NM}{{\rm NM}}
\newcommand {\SNIa}{{\rm SNIa}}
\newcommand {\tauNM}{\tau_{\NM}}
\newcommand {\tmin}{t_{\rm min}}
\newcommand {\tminNM}{t_{\rm min,NM}}

\newcommand{\mdotstar}{\dot{M}_*}
\newcommand{\mdotout}{\dot{M}_{\rm out}}

\newcommand{\taudep}{\tau_{\rm d}}

\newcommand{\taudelay}{\tau_{\rm del}}
\newcommand{\taucool}{\tau_{\rm cool}}

\title[Can NM account for r-process enrichment?]{The chemical evolution of r-process elements from neutron star mergers: the role of a 2-phase interstellar medium}
\author[R. Sch\"onrich and D. Weinberg]{Ralph A. Sch\"onrich\thanks{E-mail: ralph.schoenrich@physics.ox.ac.uk}
        and David H. Weinberg \\
        University of Oxford, Rudolf Peierls Centre for Theoretical Physics, 1 Keble Road, Oxford, OX1 3NP, UK\\
        Ohio State University, Department of Astronomy and CCAPP, 140 West 18th Avenue, Columbus OH, 43210-1173
        }

\date{Draft, \today}
\pubyear{2019}

\begin{document}
\label{firstpage}
\pagerange{\pageref{firstpage}--\pageref{lastpage}}
\maketitle

\begin{abstract}
Neutron star mergers (NM) are a plausible source of heavy r-process elements such as Europium, but previous chemical evolution models have either failed to reproduce the observed Europium trends for Milky Way thick disc stars (with $\feh \approx -1$) or have done so only by adopting unrealistically short merger timescales.  Using analytic arguments and numerical simulations, we demonstrate that models with a single-phase interstellar medium (ISM) and metallicity-independent yields cannot reproduce observations showing $\eualpha > 0$ or $\eufe > \afe$ for $\alpha$-elements such as Mg and Si. However, this problem is easily resolved if we allow for a 2-phase ISM, with hot-phase cooling times $\tau_{\rm cool}$ of order $1\Gyr$ and a larger fraction of NM yields injected directly into the cold star-forming phase relative to $\alpha$-element yields from core collapse supernovae (ccSNe).
We find good agreement with observations in models with a cold phase injection ratio $\fcNM / \fcccSN$ of order $2$, and a characteristic merger timescale $\tauNM=150\Myr$. We show that the observed super-solar $\eualpha$ at intermediate metallicities implies that a significant fraction of Eu originates from NM or another source besides ccSNe, and that these non-ccSN yields are preferentially deposited in the star-forming phase of the ISM at early times. 
\end{abstract}


\section{Introduction}
With the latest direct observation of a neutron star merger, as well as previous studies from Swift GRB counts \citep[][]{Wanderman15}, it has become fully evident that neutron star mergers (NM) are numerous enough and have a sufficient heavy element yield  \citep[][]{Smartt17} to play a significant role in shaping the chemical evolution of r-process elements like Europium (Eu). However, there has been a long tradition of chemical evolution papers challenging the impact of neutron star mergers (see below). The central issue  is the delay time distribution (DTD), i.e. the average distribution in time of events, such as Type Ia supernovae (SNeIa) or NM, after the birth of a stellar population. The delay time distribution is best characterised by an onset time ($\tmin$, short and not so important for neutron star mergers) and a characteristic timescale, e.g. the decay time $\tau$ of an exponential. Recent studies have approached the chemical evolution problem with different modelling strategies (see further below), but agree widely on the outcome: they have either assumed reasonable DTDs for neutron star mergers and concluded that they cannot explain measured abundance patterns \citep[][]{Argast04, Wanajo06}, or claimed that NM work while adopting implausibly short characteristic timescales of order $1 \Myr$ \citep[][]{Matteucci14}, which are not feasible in population synthesis.

In this paper we will show that neither argument applies. We will demonstrate that classical chemical evolution models that assume a 1-phase interstellar medium (ISM) and metallicity-independent yields cannot possibly reproduce the observed stellar abundances. Specifically, such models will always predict too low europium/r-process abundances of metal-poor stars relative to those at solar metallicity. Conversely, our models that account for differential injection into the warm/hot gas phase reproduce the high $\eufe$ ratios observed for stars with $\feh \sim -1$. Our arguments will further show that a significant fraction of r-process elements must come from a different origin than core-collapse supernovae (ccSNe).

Two concise and informative reviews of the problem of neutron star mergers and the chemical evolution of r-process elements can be found in \cite{Hotokezaka18} and \cite{Cowan19}. Evidence from the observed bolometric luminosity evolution confirms a low-$Y_e$ (electron to nucleon ratio) nucleosynthesis with sufficient heavy element production, which accounts for the origin of Lanthanides \citep[][]{Rosswog18}. Ever since the discovery of neutron stars, it has been predicted \citep[see e.g.][]{Lattimer74} that r-process elements can be generated and injected to the star-forming ISM by mergers of a neutron star with another neutron star or a black hole, when parts of the neutron star decompress rapidly and get expelled either in a tidal tail \citep[likely favouring heavier r-process nuclei, and in excellent agreement with the solar elemental composition,][]{Freiburghaus99}, or in a polar outflow \citep[likely favouring less massive nuclei,][]{Perego14, Martin15}. \cite{Thielemann17} show that the abundance pattern of r-process elements can be fully modelled by the nucleosynthesis in decompressing neutron star matter during mergers. The alternative site for r-process element generation would be pockets in ccSNe which have a large free neutron flux and low electron fraction \citep[][]{Witti94, Takahashi94}. The most promising ccSN site are collapsar models, where jets/outflows from the black hole accretion disks formed in magneto-rotational core collapse hit the stellar material \citep[][]{BK70, LeBlanc70, Nishimura17, Halevi18}. In a recent discussion, \cite{Siegel18} argue that these could be the dominant r-process production site, as their direct association with ccSNe guarantees a minimal time delay.

On the observational side, \cite{Guiglion18} show data that hint at slightly different trends of the r-process elements $\gdfe$ and $\dyfe$ vs. $\feh$ than $\eufe$, which would argue for some difference in their nucleosynthesis sources, but it remains to be seen if this trend remains robust when accounting for NLTE or 3D effects (departures from the assumed local thermodynamic equilibrium or the assumptions of 1-D atmospheres vs. a full 3D atmosphere model causing bias in the abundances inferred from stellar spectra).
While there were doubts if the rate of neutron star merger events would suffice to supply sufficient r-process elements, this is no concern any more after the recent observation of the GW170817 event by LIGO \citep[][]{Hotokezaka18, Cote18}. What remains unsolved is the central problem of the detailed enrichment history and the resulting abundance patterns.

Galactic chemical evolution models can assess if r-process elements predominantly originate in special environments in ccSNe or in neutron star mergers by exploiting their different timescales: If r-process elements form in ccSNe, there should be no difference in enrichment trends between r-process elements and alpha elements (e.g. O, Mg, Si), while r-process enrichment by neutron star mergers will show the signatures of significant time delay, i.e. under-abundant r-process elements relative to alpha elements for stars formed early in the history of the system. We will discuss the relevant DTD timescales in Section~\ref{sec:basic}. Chemical evolution studies usually use one of the following approaches: i) modelling very metal-poor stars with stochastic evolution of various flavours that model many separate domains or allow for local deviations \citep[e.g.][]{Argast04, Cescutti14, Wehmeyer15}, an approach that is vital if one attempts to fit these stars ($\feh \lesssim -2$), ii) running a classical chemical evolution model \citep[e.g.][]{Matteucci14}, i.e. one phase of ISM and one or multiple zones of the Galactic disc(s), or iii) inserting chemical evolution recipes into hydro simulations, like \cite{Voort15} or \cite{Haynes19}.

\cite{Matteucci14} formulated successful r-process enrichment models only by assuming unfeasibly short delay time distributions, with characteristic delay timescales of only $1 \Myr$. Other models, like \cite{Cescutti14} and \cite{Cescutti15}, have been quite successful in modelling halo stars with stochastic chemical evolution, but again with rather short characteristic delay timescales of $\tau_{\NM} \sim 10 \Myr$. Successful models of r-process abundances in halo stars \citep[][]{Hirai15} have been restricted to $\feh < -1$, and the continuation to higher metallicities appears problematic. On the other side of the same medal, recipes in hydro simulations, like \cite{Voort15}, get the mean $\eufe$ abundance wrong ($\eufe < 0$ instead of $\eufe >0$) along the entire metal-poor track. The real challenge (as will be shown in our discussion below) is to explain high r-process abundances in thick disc stars, while still reaching solar values at later stages, a point also noted by \cite{Cote18b}. One way out could be to go for metallicity-dependent yields, setting much larger r-process yields at low metallicity \citep[see e.g.][]{Mennekens16}, but models of this sort fit the data even worse. Solving the problem this way would also imply very high $\eufe$ values and anomalously high $\eualpha$ in dwarf galaxies, in tension with observations \citep[][]{Geisler05, Lanfranchi08}, and it would incline the $\eufe$ vs. $\feh$ plateau of the thick disc downwards (higher $\eufe$ at lower $\feh$), again in tension with the data. The collapsar model \citep[][]{Siegel18} would yield a significant change of r-process element production near solar metallicity and merits further investigation of its chemical evolution predictions, in particular a comparison with the current $\eufe(R)$ profile in the Milky Way.  

We will argue in this paper that the essential ingredient for resolving this problem is proper accounting for enrichment into different phases of the ISM.  While discussions of the ISM often distinguish three or more phases -- e.g., cold, warm, and hot -- for our purposes we need only distinguish (cold) gas that can immediately form stars from (hot) gas that must first cool on a $\sim\Gyr$ timescale before entering the star-forming phase.

Our paper is structured as follows: we start with a general discussion of the basic picture and chemical evolution models in Sections \ref{sec:metplane} and \ref{sec:basic}, followed by a description of our adopted chemical evolution model in Section \ref{sec:model}, detailing the use of different gas phases. In Section \ref{sec:analytical}, we present some general analytic arguments and constraints for chemical evolution. The models to solve the r-process abundance problem are step by step developed from a more classical chemical evolution picture in Section \ref{sec:specmodel}. In Section \ref{sec:Conclusions} we summarize the conclusions from this analysis. Appendix A1 presents an analytic model, based on the formalism of \cite{Weinberg17}, that reproduces the findings of Section \ref{sec:specmodel}. Appendix A2 demonstrates the insensitivity of our models to the assumed SNIa DTD and varies $\tau_\NM$ for a 2-phase model.

\begin{figure}
\epsfig{file=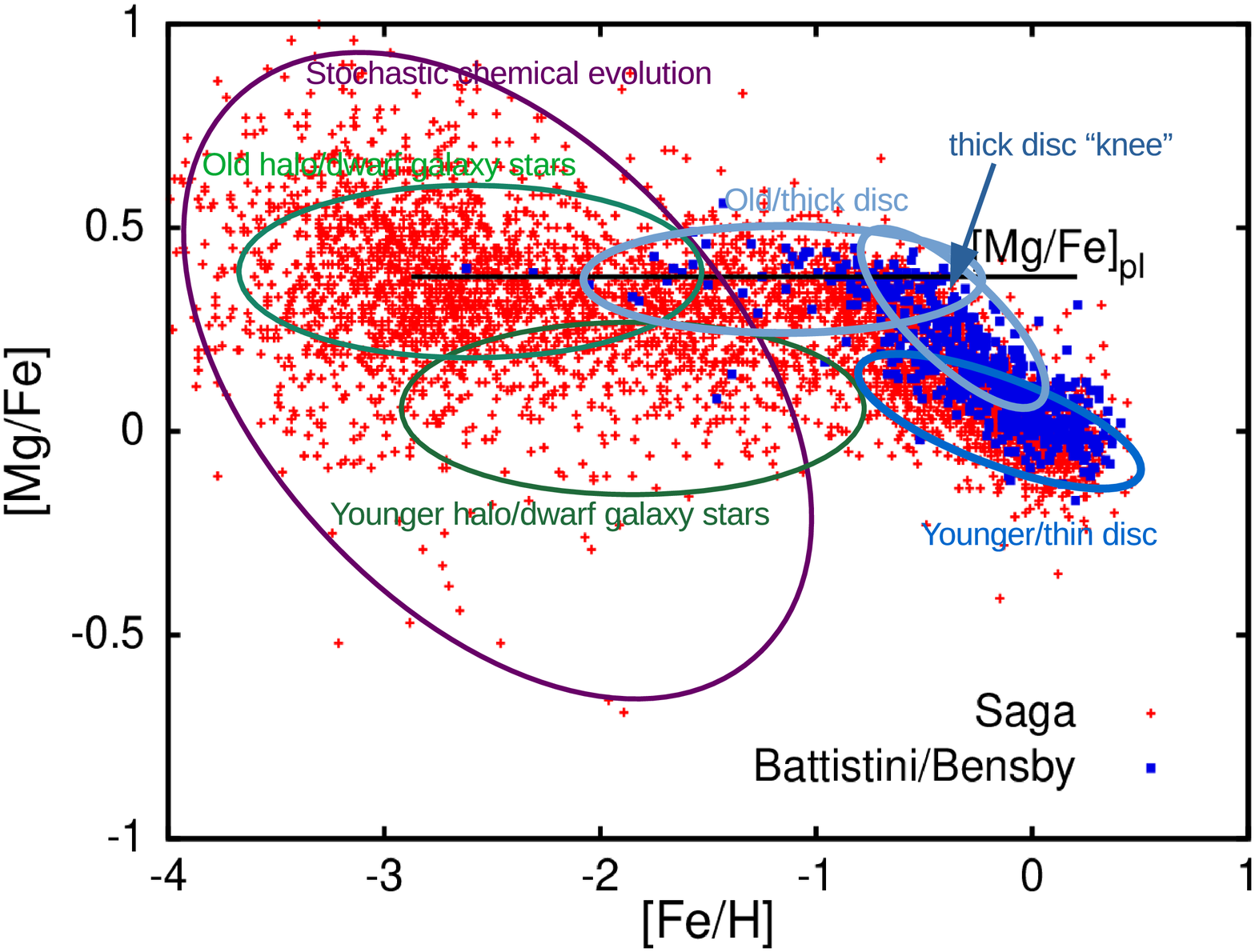,angle=0,width=\hsize}
\epsfig{file=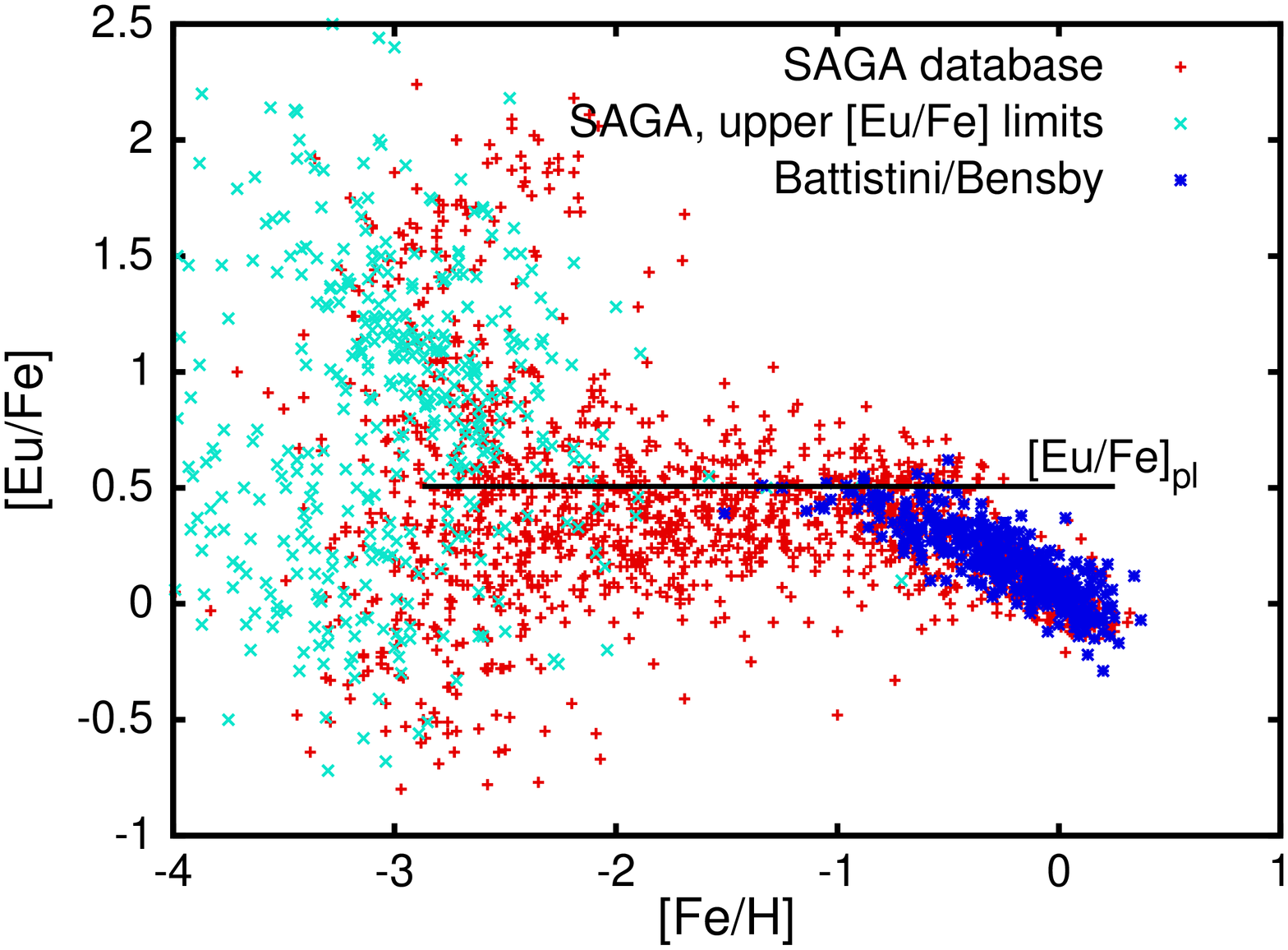,angle=-90,width=\hsize}
\caption{Top panel: Schematic diagram of the abundance plane in $\mgfe$ vs. $\feh$. The red data points show all objects in the SAGA database, while excluding all stars that have only an upper limit on either abundance. With blue data points we overplot the sample from Battistini \& Bensby, which is limited to disc stars, but offers a more homogeneous analysis and better controlled errors. We omit the error bars for each star to avoid crowding. In particular the SAGA sample is heteroskedastic, but typical precision is of order $0.1 \dex$ for abundance ratios, of order $0.07 \dex$ for single elemental abundances in the B\&B sample, with typically larger formal errors and additional systematic scatter in the SAGA database. The bottom panel shows the data for $\eu$. There are almost no stars with only upper $\eu$ limits (teal crosses) in the SAGA database at $\feh > -2 \dex$. The y-axis range in the two plots is different.}\label{fig:dataonly}
\end{figure}

\section{The abundance plane}\label{sec:metplane}

The purpose of this Section is to give a quick overview of how to interpret the structures in abundance planes, like the $\afe$-$\feh$ plane, and to introduce some of the terms used throughout this paper. Fig.~\ref{fig:dataonly} 1 displays data in the $\mgfe-\feh$ and $\eufe-\feh$ abundance planes
taken from the SAGA database \citep[][]{Suda08, Suda17} and the Bensby \& Battistini sample \citep[][]{Bensby14, BB15, BB16}. The experienced reader may wish to inspect our definition of domains in Fig.~\ref{fig:dataonly} and proceed to the next section.

Understanding the abundance planes is our only route to explore the sources of early chemical evolution. Testing characteristic delay times of order $100 \Myr$ with stellar ages, would demand systematics-proof age measurements with accuracies and precision better than $\sim 100 \Myr$ for stars that are more than $10 \Gyr$ old; a hopeless prospect at current stage, even with the accuracy of Gaia parallaxes. However, relative abundances naturally resolve this and serve as a finely resolved clock at early times: changes in relative abundances happen when new progenitor classes with different DTDs start to contribute. However, note that metallicity is usually a poor indicator of a star's age. Accreted stars from dwarf galaxies with their own clocks will overlap with low-metallicity Milky Way stars, which again have at each time a position-dependent metallicity.

Metal-poor stars are dominated by stochastic chemical evolution, i.e. Poisson noise due to a low number of contributing ccSNe, SNeIa and NM, as well as direct stochasticity, e.g. by the question where and how close a star-forming cloud is to such an event. This stochasticity drops rapidly towards larger $\feh$ (we remind that an increase by $1 \dex$ in $\feh$ equals a factor $10$ in its abundance), and becomes negligible near $\feh \gtrsim -2 \dex$. 

The features that we are most interested in are the values $\mgfe_{\rm pl}$ and $\eufe_{\rm pl}$ of the high-$\mgfe$ ridge, which is commonly the chemically defined thick disc of the Milky Way and at low $\feh$ shares its location with old halo stars. This ridge is particularly well defined in the more precise Bensby et al. data (blue points), as well as the location of the thick disc knee, which marks the set-in of SNeIa in the inner Galactic disc, and which is nearly guaranteed to be free of any contamination by halo stars or stochastic chemical evolution effects. Lower $\mgfe$ values imply that SNeIa have contributed a significant fraction of the iron budget (i.e. at least a few Gyrs after the formation of the system) and thus designate younger objects: the thin disc on the high $\feh$ end separated from the younger halo star populations at $\feh \lesssim -0.7 \dex$. 
We note that this $\afe$ clock is again not an absolute time - systems have different star formation histories, and $\afe$ values may depend on galactocentric radius even in the Milky Way.

\begin{figure}
\epsfig{file=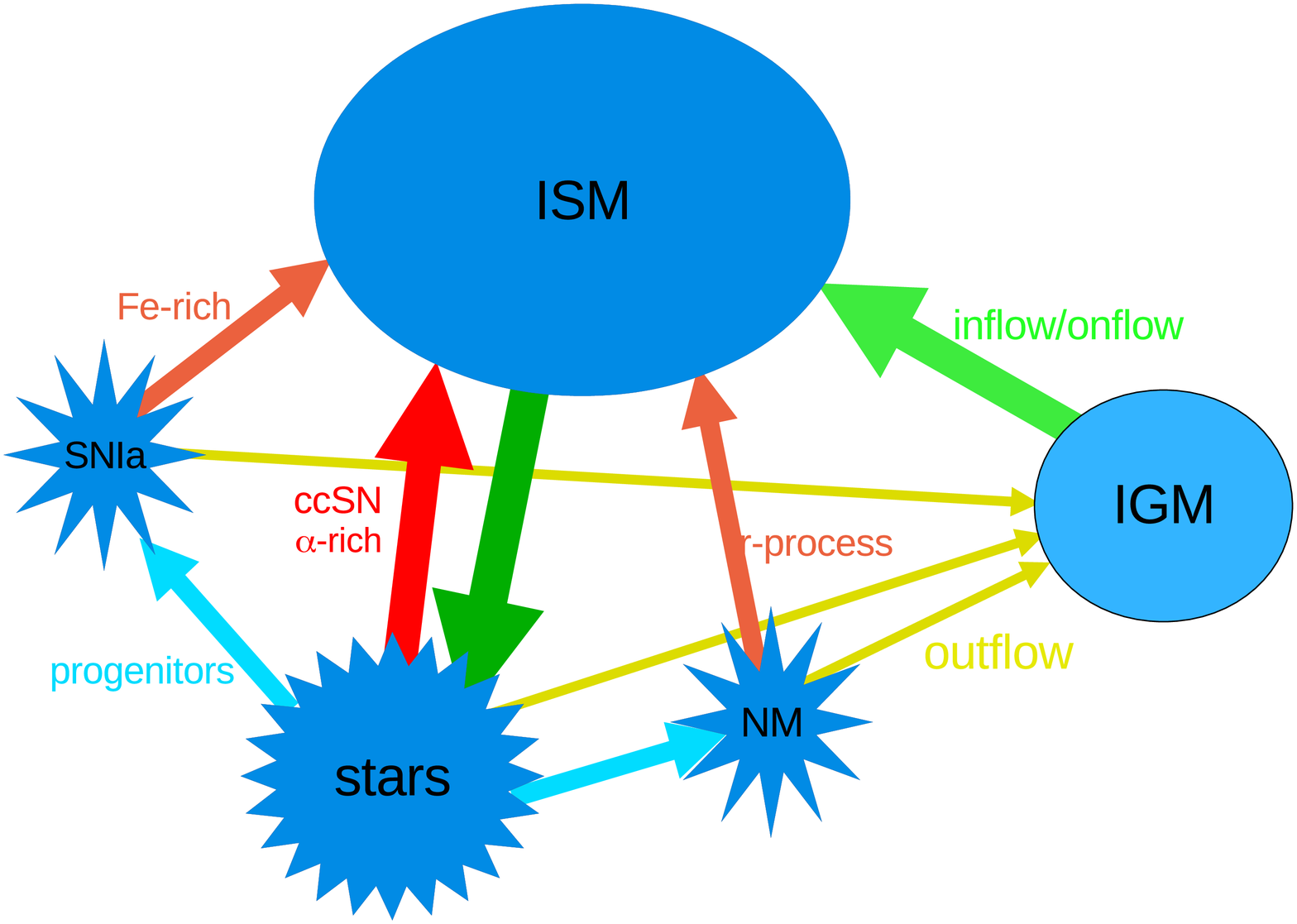,angle=0,width=\hsize}
\epsfig{file=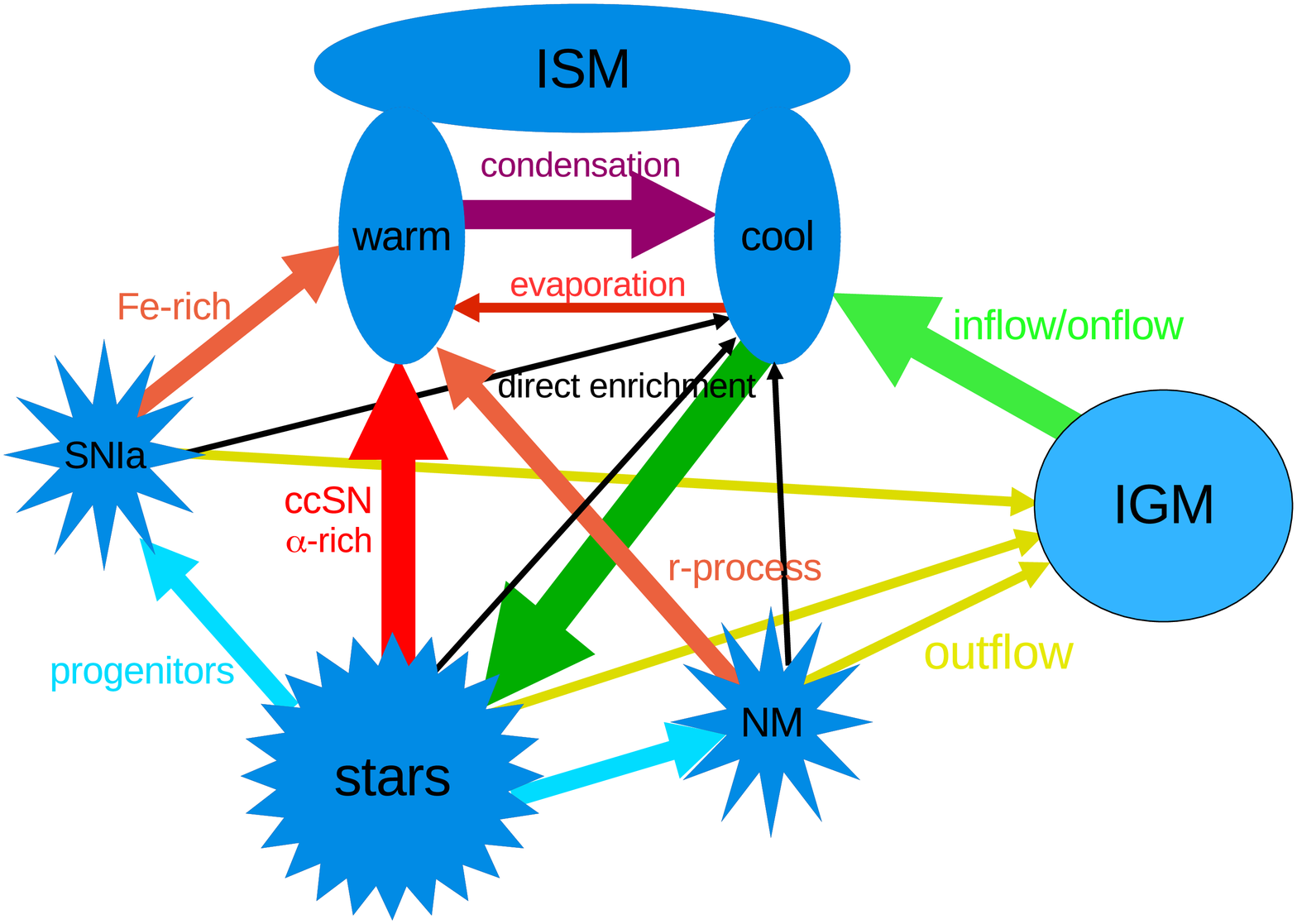,angle=0,width=\hsize}
\caption{Schematic picture of the gas balances in a chemical evolution with a 1-phase ISM (top) and a 2-phase ISM (bottom). In the latter case, we will express the fraction of yields going directly to the cold gas phase (direct enrichment) with $f_c$. To sustain star formation, any galaxy like the Milky Way must accrete significant amounts from the inter-galactic medium (IGM), and the usual assumptions for yields combined with a Kroupa/Chabrier IMF compared to observed thin disc metallicities imply that a bit more than half of a galaxy's elemental yields should be lost. This loss may also carry gas from the ISM to the IGM, a channel not marked here.}\label{fig:modelbasics}
\end{figure}

\section{Basic chemical evolution: sourcing the elements}\label{sec:basic}

To understand the chemical evolution processes shaping the distribution of stars in abundance space, it is useful to draw the gas balances. Fig.~\ref{fig:modelbasics} shows a graph of gas balances in a 1-phase (top) vs. a 2-phase model accounting for a hot ISM (bottom). To understand the problem of r-process abundances, we only need to account for the three progenitor classes that produce iron (Fe), alpha elements, and r-process elements: ccSNe, SNeIa, and NM. Each of these classes will, with its own delay time distribution, yield its own mix of elements back to the ISM, from where new stars are formed. 

The functional shape of DTDs is frequently taken to be $t^{-1}$ for SNeIa; this shape is suggested by population synthesis models compared with data \citep[][]{Hachisu08, Totani08}. Similarly, the DTDs for neutron star mergers is frequently described by a $t^{-1}$ or $t^{-1.5}$ law. From a practical point of view, this functional shape is unattractive, since it is not integrable to at least one side, thus requiring a cut-off and normalisation conditions sensitive to this cut-off. Here, we will adopt a simple exponential: ${\rm DTD}(t) = \exp(-t/\tau)$, where $\tau$ is the characteristic timescale. In Appendix A2 we show that using a double exponential law for SNeIa, which fits the power law more closely, does not affect our results. The difference for NM is even less significant due to the shorter timescale.

Let us outline these three progenitor classes:
\begin{itemize}
 \item Massive stars exploding (usually as ccSNe): most massive stars attain an onion-like structure just prior to collapse, with an Fe/Ni core at the centre, surrounded by shells dominated by $\alpha$ elements of decreasing mass number. In the explosion, most of the shells are expelled into the ISM, providing the bulk of today's inventory of $\alpha$ elements, like Si, or Mg, or Ca. The mass cut where material can still escape the emerging neutron star determines the early production of Fe. Theoretical models of ccSNe do not predict the mass cut robustly. Some empirical indications come from estimates of $^{56}{\rm Ni}$ mass in ccSNe. The picture of a simple mass cut is only a naive approximation to real explosions, where convection and anisotropy (evidenced by neutron star kicks and remnant observations) allow for some expulsion of material below the average mass cut. However, the average $\fe$ yield from ccSNe is constrained by the $\afe_{\rm pl}$ value of the thick disc (see Fig.~\ref{fig:dataonly}). The value of $\afe_{\rm pl}$ is close to $\log_{10}{2} = 0.3$, which shows that ccSNe account for roughly half of today's $\fe$ inventory. The massive progenitors of ccSNe ($M \gtrsim 10 \Msun$) explode after lifetimes of a few Myrs to less than $100 \Myr$. 
 \item Exploding white dwarfs (SNeIa): we know that the remaining half of today's $\fe$ must be produced in SNIa explosions. There is still considerable debate about which systems account for which fraction of SNeIa, e.g. accretion onto a white dwarf to reach the Chandrasekhar mass $\Mch$, or triggered explosions below $\Mch$ by a pressure wave from surface He burning, or white dwarf mergers in a double degenerate system. The common denominator is the explosion of a C/O white dwarf, which will burn about half of its mass directly to $^{56}{\rm Ni}$ decaying into Fe. We use a characteristic delay timescale of $\tau_\SNIa = 1.5 \Gyr$.
 \item The source of r-process elements: either neutron star mergers or pockets with particular conditions (jets, neutrino driven winds, collapsar scenario) in ccSNe. In the latter case, the delay time distribution is nearly identical with ccSNe. For NM the DTD timescale should be faster than that of SNeIa, since the progenitor systems are more massive, arise on the ccSN timescale, and at least the single degenerate scenario for SNeIa depends on the evolution of the secondary star. Also, the Swift gamma ray burst counts show that $\sim 40\%$ of the neutron star mergers have no measurable delay relative to star formation \citep[][]{Wanderman15, Fong17}. The frequently used $t^{-1}$ or $t^{-1.5}$ DTD \citep[][]{Hotokezaka18, Cote18} combined with their shorter minimum delay time relative to SNeIa implies an equivalent characteristic timescale $\tauNM$ that is also about an order of magnitude shorter than for SNeIa. We therefore take $150 \Myr$ as a best guess for $\tauNM$, but we will investigate a range of values.
\end{itemize}

Fig.~\ref{fig:modelnohot} shows the tracks of cold ISM abundance through the $\mgfe$ vs. $\feh$ abundance plane. All models (described further in Section~\ref{sec:model}) start near metal-free, and move from left to right. The initial abundance ratios (y-axis) on the left-hand side are always set by the fastest channel of enrichment. At $\feh > -2.5 \dex$, they are set by all sources that contribute on a timescale of about $100 \Myr$. The time spacing between model points at each of the three plotted galactocentric radii is $15 \Myr$. The onset of SNIa explosions begins to drive down $\eufe$ and $\mgfe$ around $\feh \gtrsim -1 \dex$. We will later see that accounting for a hot phase, which can lock up yields for of order $1 \Gyr$, can significantly alter the effective DTD for yields from the two faster progenitor classes (ccSNe, NM), or even reverse their temporal order, if their cold channel fractions $f_{c,i}$ differ.

\section{Galaxy model}\label{sec:model}

\subsection{General description}

For this paper we use the \cite{SB09a} model with the parameters of \cite{SM17}. A full description is provided in those two papers, so we will limit ourselves to just some basic facts. Similar to classical chemical evolution models \cite[][]{Chiappini97} the model divides the galactic disc into concentric rings, with a spacing of $\sim 0.2 \kpc$, using a timestep of $15 \Myr$. We experimented with higher resolutions, but found that they do not matter for modelling chemical evolution in general, as long as we stay out of the extremely metal-poor regime, where results are anyway dominated by stochastic chemical evolution. The model fully accounts for time-dependent yields from stars, radial migration in both stars and gas, and stellar orbital motions (which further broaden the distribution in stellar radii beyond the radial migration). The model has an inflow of fresh gas to sustain star formation \citep[][]{Chiosi80}, which by angular momentum conservation drives radial flows \citep[][]{Lacey85} into the disc according to the prescription of \cite{Bilitewski12}. The relative angular momentum of infall (about $0.75$ of the disc angular momentum) is fitted to match the radial abundance profile of the Galactic disc as measured with Cepheids.

Yields for Mg, Si, and Fe in the ccSNe are taken from \cite{Chieffi04}, and we use the SNIa yields for all three elements from \cite{Iwamoto99}. The results are compared to the solar abundance scale from \cite{Grevesse07}. The solar abundances for Mg and Fe are well-measured. One can argue a little about the yields, but ultimately this comes down to fixing the iron yield from ccSNe to fit the thick disc plateau and then adapting the SNIa rates to reach solar abundances; thus, using different sources would not significantly affect our conclusions. The yields for r-process elements are still very uncertain, and attempts to compile such data would only distract from our main aim to investigate the range of possible delay-time distributions. Thus, we have implemented r-process yields as a generic element and normalise them such that at late times the standard model achieves $\rfe = 0$. 

An open question is the loss rate of metals to the IGM. If one assumes a Salpeter IMF, only minimal losses of order $10 \%$ are needed to reach near-solar metallicity in the local thin disc. However, the now favoured Chabrier or Chabrier-like (e.g. Kroupa) IMF has far less low-mass dwarf stars to lock up metals, and so implies a higher loss-rate of $\sim 0.6$ (only $40 \%$ of metals are retained). Yields from stellar populations and thus required loss rates could also be lower if many massive stars collapse directly into black holes without a significant explosion \citep[][]{Adams17}. 

In the numerical chemical evolution models applied here, metal loss is treated by a direct loss from the galaxy without evaporating a significant amount of the ambient cold ISM. The opposite extreme is ejection of gas at the ISM metallicity characterised by a mass loading parameter $\eta = \dot{M}_{\rm out} / \dot{M}_{\star}$. In the prescription of \cite{Andrews17}, which does not implement radial flows and thus needs a higher loss rate, we require $\eta = 2.5$. We show in Appendix A1 that the choice of prescription affects the chemical evolution of $\eufe$ and $\feh$ at very low metallicity, $\feh \lesssim -2 \dex$, but does not affect our conclusions based on the thick disc metallicity regime. The SNIa rates in the model have been adjusted to match the $\mgfe$ plane, and are consistent with expectations for the Milky Way. As discussed in \S 5, we believe that $\mgfe$ ratios may be artificially boosted by NLTE effects, making Si a more reliable reference element in measurements. However, our SNIa rates are calibrated to match observed $\mgfe$ rather than $\sife$, and rather than readjust the rates when comparing to Si observations, we simply reduce model $\si$ values by $0.07 \dex$ (and thus decrease $\sife$ and increase $\eusi$ by the same amount).

The delay time distribution for SNeIa is set as a single exponential with timescale $\tau_{\rm SNIa} = 1.5 \Gyr$, with a minimum delay time of $150 \Myr$. We show in Appendix A2 that the precise formulation of this law does not affect our results significantly. Analogously, we use a minimum time delay of $15 \Myr$ and several choices of timescale $\tau_{\NM}$ for neutron star mergers.

\subsection{Distribution of elements to different gas phases}\label{sec:diffgasphases}

The main benefit of using the \cite{SB09a} model is that it accounts with a simple 2-phase ISM for the presence of hot gas, which cannot directly partake in the star formation. But which elements and which yields go where? One could be tempted to think that all yields are first locked up in the hot gas phase, and then cool back into the star-forming ISM. On the timescales we are interested in, however, a few observations show that this is not the case; for example, the short-lived isotopes identified in Earth's crust and solar system meteorites \cite[][]{Wasserburg06} prove that there must be a direct channel from stellar yields to the cold star-forming ISM. Further, we know for sure that at least the ejecta from novae \citep[][]{Amari01} and even from ccSNe \citep[e.g.][]{Travaglio99} can directly form dust grains, which made their way directly to the meteorite formation sites of the nascent solar system. 

Due to their long delay time, SNeIa will commonly explode far from the star forming regions, so that their ejecta are not trapped by a dense surrounding ISM, and they are also more energetic than ccSNe. Neutron star mergers will be different from ccSNe, and are likely affected by the different timing: the early events should still take place predominantly in dense gas regions, but the later ones might not. Most importantly, NM produce relatively dense ejecta with very heavy chemical elements. This matters, since a \enquote{normal} ISM can achieve the well-known metastable hot state, where a few million Kelvin hot plamsa is fully ionised and thus has much lower cooling rates than at, say, $10^5 \K$. However, the ionisation temperature goes with the first ionisation energy, which in turn depends on $Z^2$. So, while hydrogen has a typical ionisation temperature of $10^4 \K$, this temperature is of order $10^8 \K$ for uranium, i.e. there is basically no astrophysical system where heavy r-process elements are fully ionised, not even when the near-relativistic ejecta shock with the surrounding ISM. Therefore, we expect a larger fraction of NM yields to cool quickly.

Given the above, we can make the following assumptions
\begin{itemize}
 \item We put the bulk of SNIa yields ($0.99$) into the hot gas phase, i.e. $\fcSNIa = 0.01$. The choice hardly affects chemical evolution (see below).
 \item About $\fcccSN = 0.25$ of ccSN yields is fed to the cold gas phase. We test different values.
 \item The fraction of neutron star merger material directly fed the cold phase is set as $\fcNM = \fcccSN$ for our initial models and later varied independently.
\end{itemize}

We note that $\fcSNIa$ is not important to our discussion, since the cooling timescale is still even a bit shorter than the characteristic SNIa delay time, i.e. increasing $\fcSNIa$ would not significantly alter our results. The cooling timescale (again exponential) is set to $1 \Gyr$ as in SB09, in concordance with the mass of warm/hot ISM found in the Milky Way.

\section{Some analytical considerations about abundance levels}\label{sec:analytical}

The key observational results that will drive our conclusions in this 
paper are that disk stars in the metallicity range  $\feh \sim -1.5$ to $-0.5$ have typical $\eumg > 0$, typical $\eusi \sim 0.1-0.2$, and typical $\eufe \ga \mgfe$ (see Figures~3 and 4 below).

Before proceeding to detailed comparison of numerical models to observations, it is useful to derive some basic expectations and implications from analytic arguments.  Here we focus on limiting cases, with Appendix A1 presenting a more complete analytic model that confirms these limits and supports our numerical findings.

We begin by making the conventional assumption of a 1-phase, well mixed ISM at all times.  If the IMF averaged ccSN yields of Mg and Fe are independent of metallicity, then at early times before SNIa enrichment is important the ratio of these elements in the ISM must simply equal the ratio of their ccSN yields, implying a plateau ratio
\begin{equation} 
\mgfe_{\rm pl} = \log{\frac{y_{\mg,\ccSN}}{y_{\fe,\ccSN}}} -
\log{\frac{M_{\mg,\odot}}{M_{\fe,\odot}}} .
\end{equation}
The flatness of the observed $\mgfe$ plateau supports the assumption of metallicity independent yields, which is also consistent with theoretical expectations for $\alpha$ elements. At later times, SNIa enrichment adds Fe without associated Mg, thus depressing $\mgfe$.  The solar ratio $\mgfe=0$ reflects this latter condition, with roughly equal Fe contributions from ccSNe and SNeIa. As a rough approximation, we expect
\begin{equation}
 \mgfe_{\rm pl} \sim \log{\left(
  \frac{y_{\fe,\ccSN} + y_{\fe,\snIa}}{y_{\fe,\snIa}}\right)}
  = \log{\left(1 + \frac{y_{\fe,\snIa}}{y_{\fe,\ccSN}}\right)} ~,
\end{equation}
though in detail the drop from the plateau to the late-time equilibrium
ratio depends on the star formation history \citep[see, e.g., figure~3 of][]{Weinberg17}.

Now let us assume that the $r$-process yield from NM is also independent
of metallicity, and that the time delay for NM is much shorter than
for SNeIa.  In this case the Eu/Mg rises quickly from zero to a ratio
that reflects the ratio of yields, from NM and ccSNe, respectively.
Provided the yields remain metallicity independent up to solar $\feh$,
this ratio must correspond to the solar value $\eumg=0$.  Since
$\eufe = \eumg + \mgfe$, we have
\begin{equation}
\eufe_{\rm pl} \leq \mgfe_{\rm pl} $,$
\end{equation}
which is an inequality because Eu production always lags at least
slightly behind Mg production. If the NM delay time is not sufficiently
short, then SNIa enrichment can depress $\mgfe$ before $\eumg$ rises
to the yield ratio, in which case there is not a flat plateau of $\eumg$,
but at all times
\begin{equation}
\eufe \leq \mgfe~.
\end{equation}
The above inequalities approach equality for very short NM characteristic delay times. 
If the dominant source of Eu is ccSNe rather than NM, and yields
are metallicity independent, then we simply have $\eumg=0$ and
$\eufe=\mgfe$ at all times. Yields of alpha elements from SNeIa allow for a small violation of the above inequalities (see below).

These arguments tell us that a 1-phase model with metallicity independent
yields cannot explain observations showing $\eumg_{\rm pl} > 0$ or
$\eufe_{\rm pl} > \mgfe_{\rm pl}$, regardless of whether ccSNe or
NM are the main source of Eu. 

Spectroscopic measurements of $\mgfe$ are subject to significant non-LTE corrections \citep[][]{Bergemann17}, which become stronger towards lower metallicity. We thus feel compelled to compare to Si, as a second alpha element, which should have different biases. However, Si likely has a much stronger yield contribution from SNeIa. This effect is anticipated in theoretical yield models from \cite{Iwamoto99}, which have an $\fe / \mg$ compared to the Sun that is an order of magnitude larger than for $\fe /\si$. The resulting change between $\mg$ and $\si$ in the SM17 chemical evolution model still underpredicts what is inferred from empirically calibrated yield models \citep[][]{Rybizki17} and from $\simg$ ratios in APOGEE \citep[][]{Weinberg18}, which both indicate that $\sim 20 \%$ of the solar $\si$ content are contributed by SNeIa. A contribution of this magnitude allows a violation of the above inequalities by a factor $1.25$ or about $0.1 \dex$, i.e.
\begin{equation}
\eusi_{\rm pl} \leq [{\rm Eu}/\si_\ccSN] + 0.1
\end{equation}
and
\begin{equation}
\eufe_{\rm pl} \leq \sife_{\rm pl} + 0.1 $.$
\end{equation}
For the yields in our numerical model, the allowed violation is only about $0.05 \dex$.

Things change if we allow for a hot ISM phase and assume that only a fraction $\fcccSN$ of ccSN yields and $\fcNM$ of NM yields go directly into the cold, star-forming ISM.  Before there is time for cooling from the hot phase, this model has the same effect as reducing yields by a factor $f_c$.  However, these factors do not enter the normalization to solar abundance ratios provided the ISM cooling timescale is short relative to the time needed to evolve to solar $\feh$.  Our previous results therefore adjust to
\begin{equation}\label{eq:eufe2phase}
\eumg_{\rm pl} \leq \log_{10}\frac{\fcNM}{\fcccSN}
\end{equation}
and
\begin{equation}
\eufe_{\rm pl} \leq \mgfe_{\rm pl} + \log_{\rm 10}\frac{\fcNM}{\fcccSN}~.
\end{equation}
For Si in place of Mg, the same inequalities hold but with $+0.1$ on the right hand sides.  We thus see that the observational constraints can potentially be satisfied if $\fcNM > \fcccSN$, as we have argued is plausible in Section~\ref{sec:diffgasphases}.  For example, with $\fcccSN = 0.25$ and $\fcNM = 0.5$, values of $\eufe$ can rise $\sim 0.3 \dex$ over the $\mgfe$ plateau or $\sim 0.4 \dex$ over the $\sife$ plateau.

One could try to employ 1-phase models to explain the observed Eu evolution by invoking enhanced production at low metallicity, thus boosting $\eumg$ and $\eufe$ in the plateau regime relative to solar abundance. Metallicity dependence of NM yields could arise from changing binary populations or neutron star production rates, as suggested by \cite{Mennekens16}. However, their resulting chemical evolution models do not match the observations they compared to. In general, such a model would at least require fine-tuning, and potentially impossible parameter choices; it naturally predicts enhanced $\eufe$ in low  metallicity dwarf galaxies, which does not appear supported by observations \citep[][]{Geisler05, Lanfranchi08}, and it appears in contrast to the apparent flatness of  $\eufe_{\rm pl}$ against $\feh$ in the SAGA database at low metallicities (see Fig.~\ref{fig:dataonly}). By contrast, we will show that the 2-phase ISM reproduces observed tracks with plausible ISM physics and parameter choices.

One could mimic the results of our 2-phase model by invoking differential outflow efficiencies for ccSN and NM products at early times that go away at late times, since this effectively changes the relative $\alpha$-element and $r$-process yields between sub-solar and solar metallicities.  However, this scenario again requires contrived parameter tuning relative to the natural results of the 2-phase model.

\begin{figure}
\epsfig{file=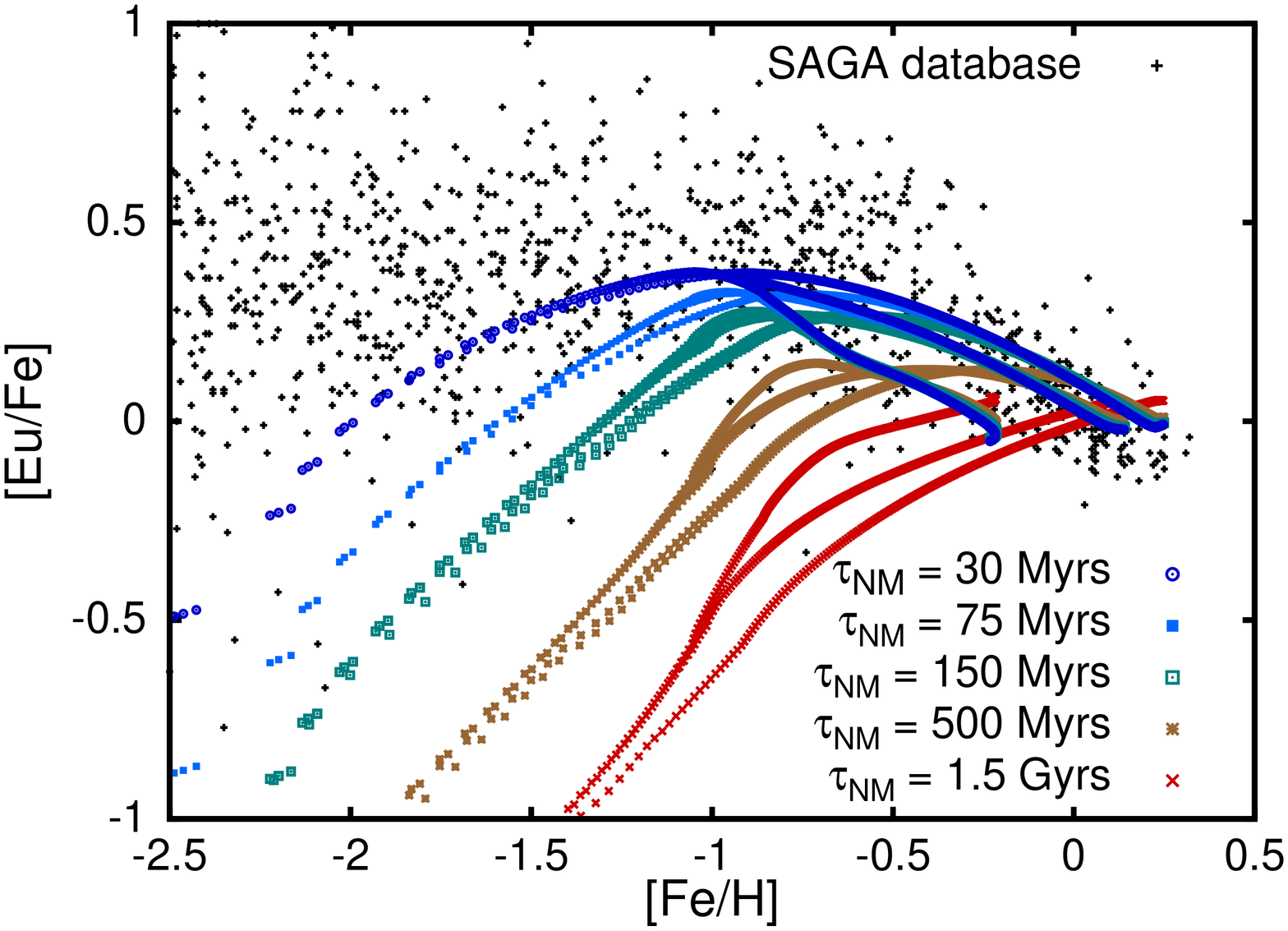,angle=-90,width=\hsize}
\epsfig{file=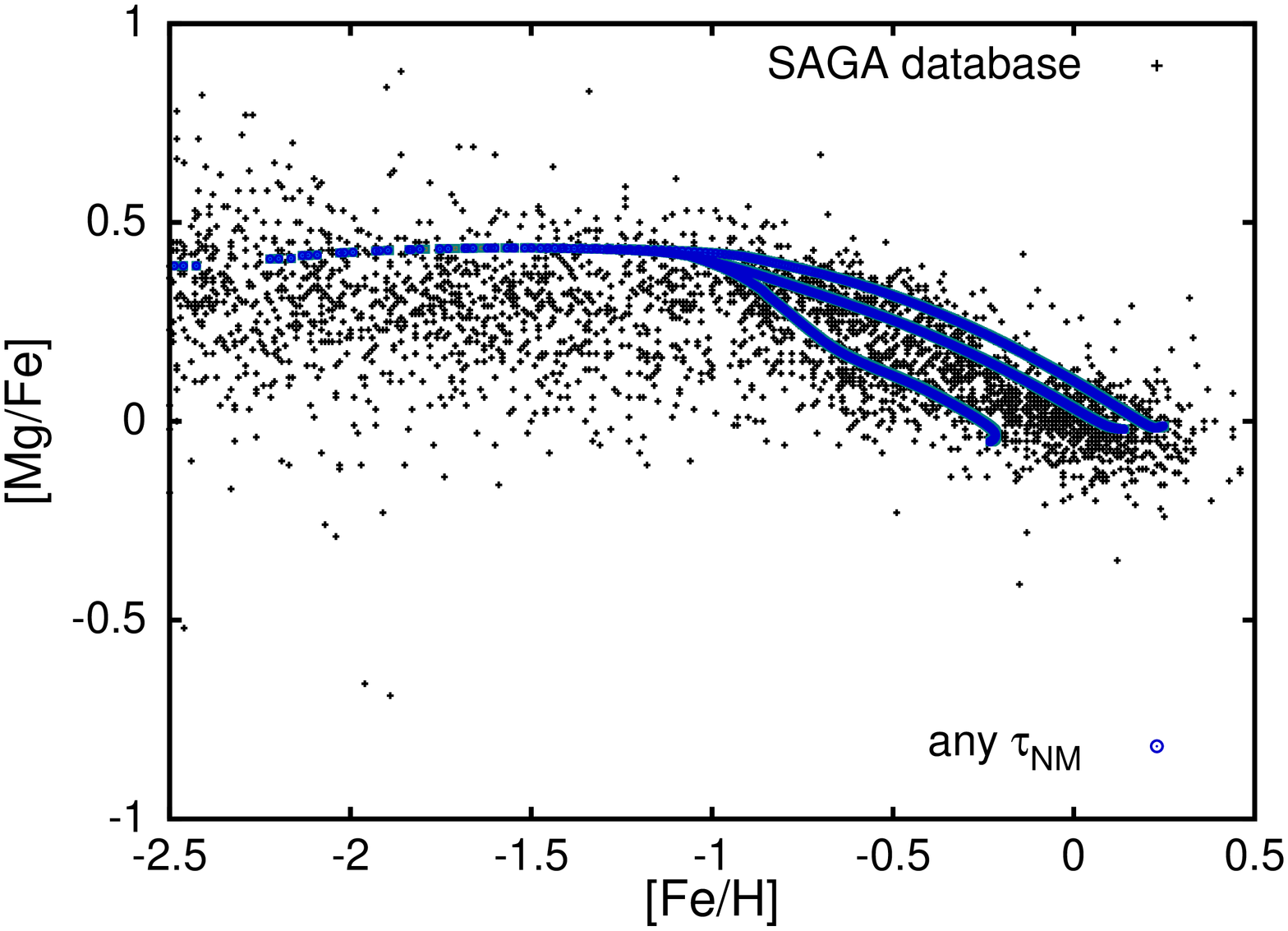,angle=-90,width=\hsize}
\caption{Model trajectories at galactocentric radii $R = 4,6,10 \kpc$ for $\eufe$ (top panel) and $\mgfe$ (bottom panel) vs. $\feh$. The crosses indicate positions of stars in the SAGA database. These models all use a 1-phase ISM and the single exponential decay timescale prescription for SNeIa. The different colours encode models with different exponential decay timescales for neutron star mergers, $\tau_{\rm NM}$.}\label{fig:modelnohot}
\end{figure}

\begin{figure*}
\epsfig{file=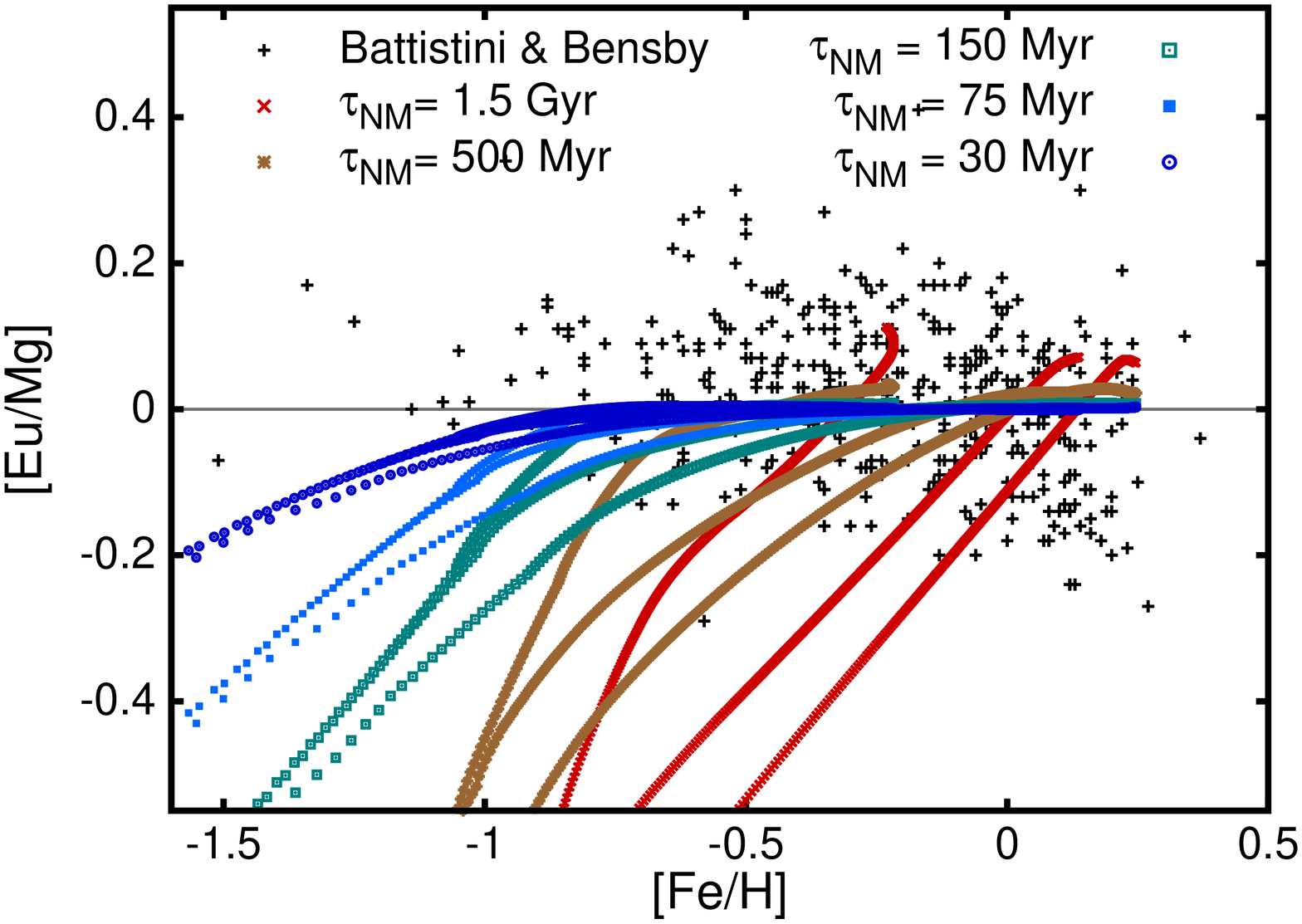,angle=-90,width=0.49\hsize}
\epsfig{file=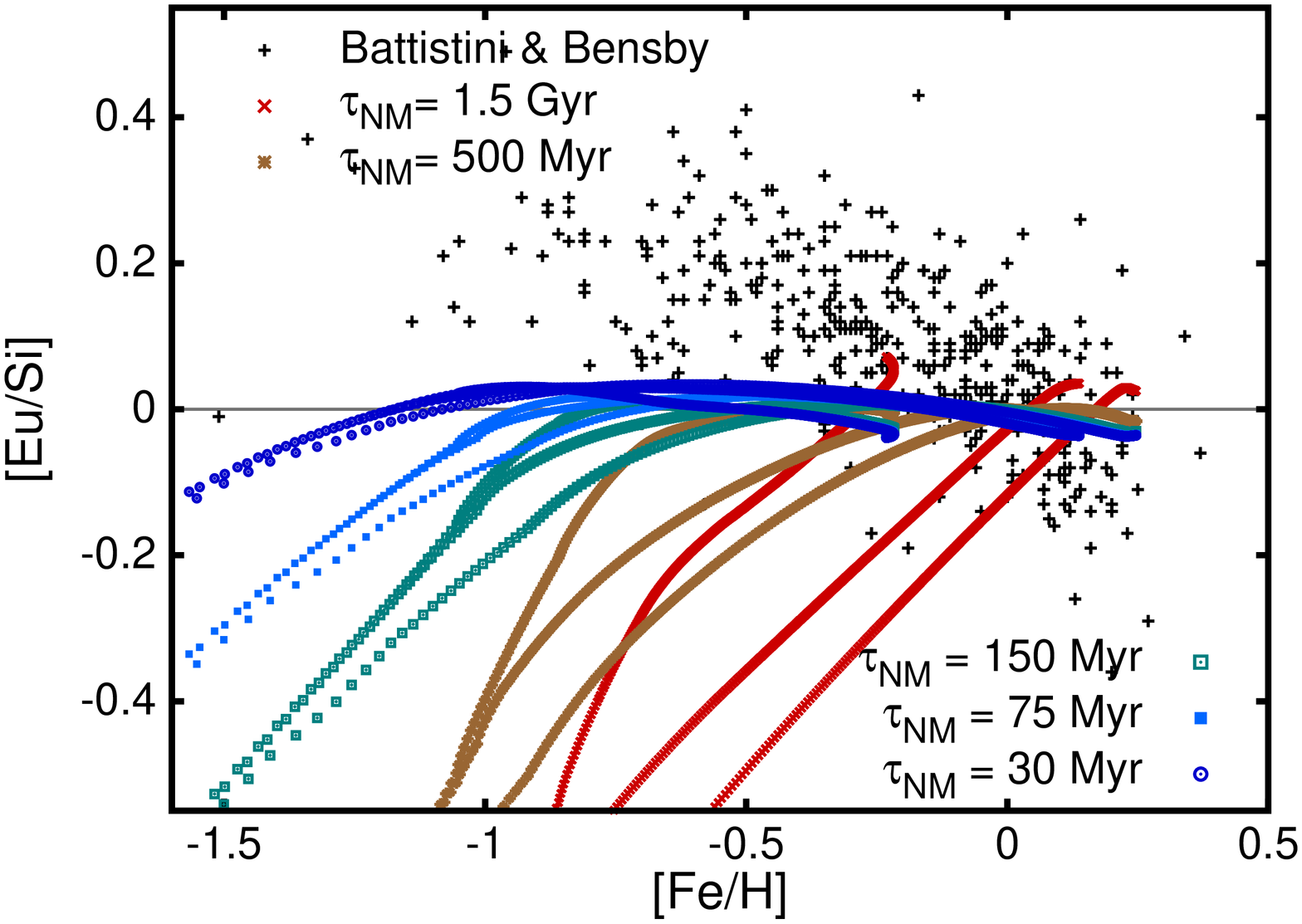,angle=-90,width=0.49\hsize}
\epsfig{file=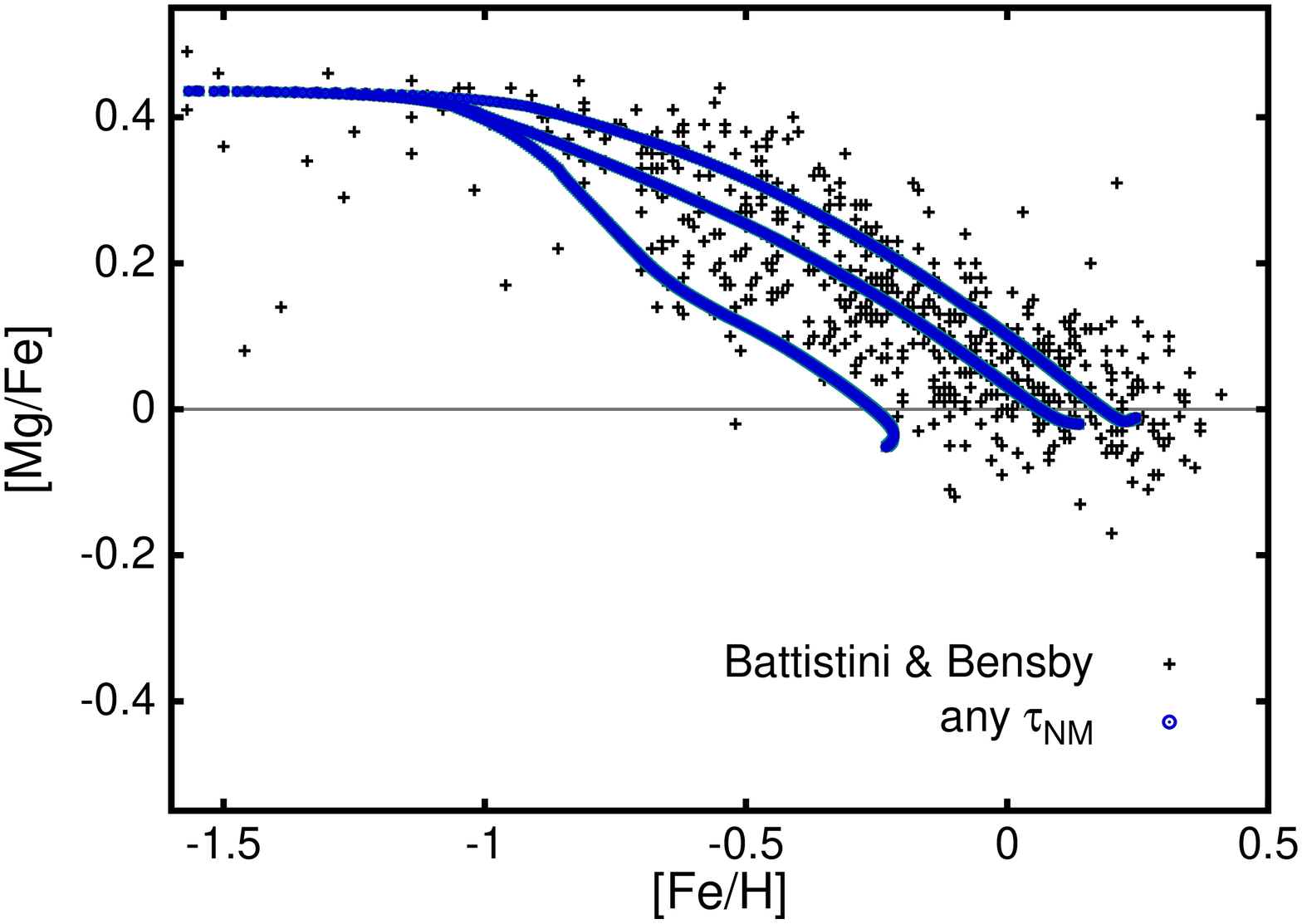,angle=-90,width=0.49\hsize}
\epsfig{file=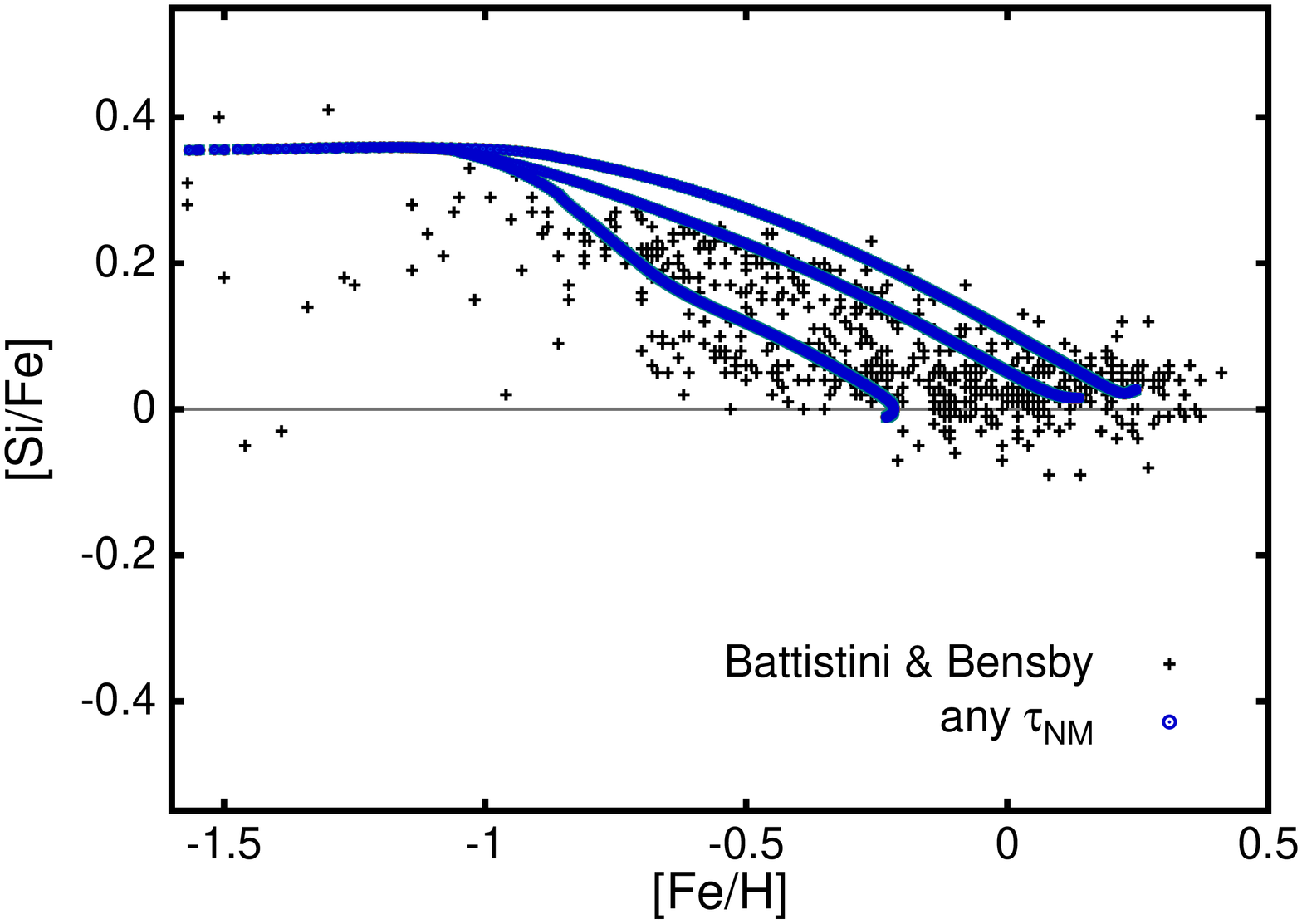,angle=-90,width=0.49\hsize}
\caption{Model trajectories as in Fig.~\ref{fig:modelnohot}, now overplotted on data from Battistini \& Bensby (2016) and Bensby et al. (2014). We now show against $\feh$ on the x-axis the discrepancy between models and data for $\eumg$ (top left panel) and $\eusi$ (top right). As shown in Section~\ref{sec:analytical}, 1-phase models are bound to stay below $\eualpha \lesssim 0$ (the slightly positive model $\sife$ values in the top right panel derive from the SNIa $\si$ yield), while the data clearly show $\eualpha > 0$ in particular for thick disc stars. Note that non-LTE analysis \citep[][]{Bergemann17} shows that observed thick disc $\mgfe$ are overestimated and $\eumg$ are underestimated by $\sim 0.1 \dex$.}\label{fig:modelMgSinohot}
\end{figure*}

\section{Models and comparisons in abundance space}\label{sec:specmodel}

After these general discussions we perform some numerical experiments using the \cite{SM17} model and overplotting those on the collected data in the SAGA database \citep[][]{Suda08, Suda17}. The bottom panel of Fig.~\ref{fig:modelnohot} shows the evolution of $\mgfe$ vs. $\feh$ for three different radii ($4,6,10 \kpc$) in the chemodynamical model. For these first plots, we have removed the hot gas phase, i.e. all stellar yields are either lost to the IGM or directly enter the star-forming gas. The evolution follows the usual picture: all trajectories run from left to right at $\mgfe_{\rm pl}$, forming the thick disc ridge. Once the SNeIa set in, the trajectories tend towards solar $\mgfe$, while their different positions in $\feh$ reflect the Galactic radial metallicity gradient. The top panel shows the modelled r-process abundance from neutron star mergers using different timescales vs. the $\eufe$ observations compiled in the SAGA database. The qualitative picture is straight-forwardly explained. The initial rise in $\rfe$ derives from the relative time-delay of NM vs. ccSNe. Trajectories can reach a high $\rfe$ plateau when both i) $\tau_\NM \ll \tau_\SF$, where $\tau_\SF = \left(d\ln(\rm SFR)/dt\right)^{-1}$, i.e. the star formation rate must not rise too quickly, ii) $\tau_\NM \ll \langle \tau_{\rm pop} \rangle$, where $\langle \tau_{\rm pop}\rangle$ is the average population age, and iii) $\tau_\NM \ll \tau_\SNIa$, i.e. the SNeIa do not significantly contribute on the NM timescale. Thus, only the shorter timescales with $\tau_\NM < 150 \Myr$ can form a high $\rfe$ plateau. In accordance with our approximations in Section~\ref{sec:analytical}, no matter how short we choose $\tau_\NM$, $\eufe_{\rm pl}$ just approaches the value of $\mgfe_{\rm pl}$, but does not reach or exceed it. We will in later figures fix $\tau_\NM = 150 \Myr$, but this choice has little impact on our results as long as we keep $\tau_\NM \ll \tau_\SNIa$.

The tension between 1-phase models and data becomes even more evident in Fig.~\ref{fig:modelMgSinohot}, which shows the behaviour of $\eumg$ (top left) and $\eusi$ (top right) vs. $\feh$. For assessment of the calibration we show the respective $\afe$ vs. $\feh$ trends in the bottom row. Here we use the merged catalogue from \cite{Bensby14}, \cite{BB15} and \cite{BB16}, so that we have only data from one single, consistently analysed disc sample. While this does not remove all systematic problems, it reduces possible trends arising from heterogeneous data sets. In the lower panels $\sife_{\rm pl}$ shows the slightly lower value expected for an element that has a slight production also from SNeIa. As argued in Section~\ref{sec:analytical}, the model $\eumg$ values never rise significantly above $\eumg > 0$, nor above $\eusi > 0.05$, where the slight violation is permitted by the small SNIa contribution to Si yields. This is in stark contradiction to the data, which feature $\eusi \gg 0.05 \dex$ and $\eumg \gg 0 \dex$, despite the fact that non-LTE analysis \citep[][]{Bergemann17} shows that in the data presented here, $\mgfe_{\rm pl}$ and consequently the observed $\eumg$ values are biased by $\sim -0.1 \dex$.

\begin{figure}
\epsfig{file=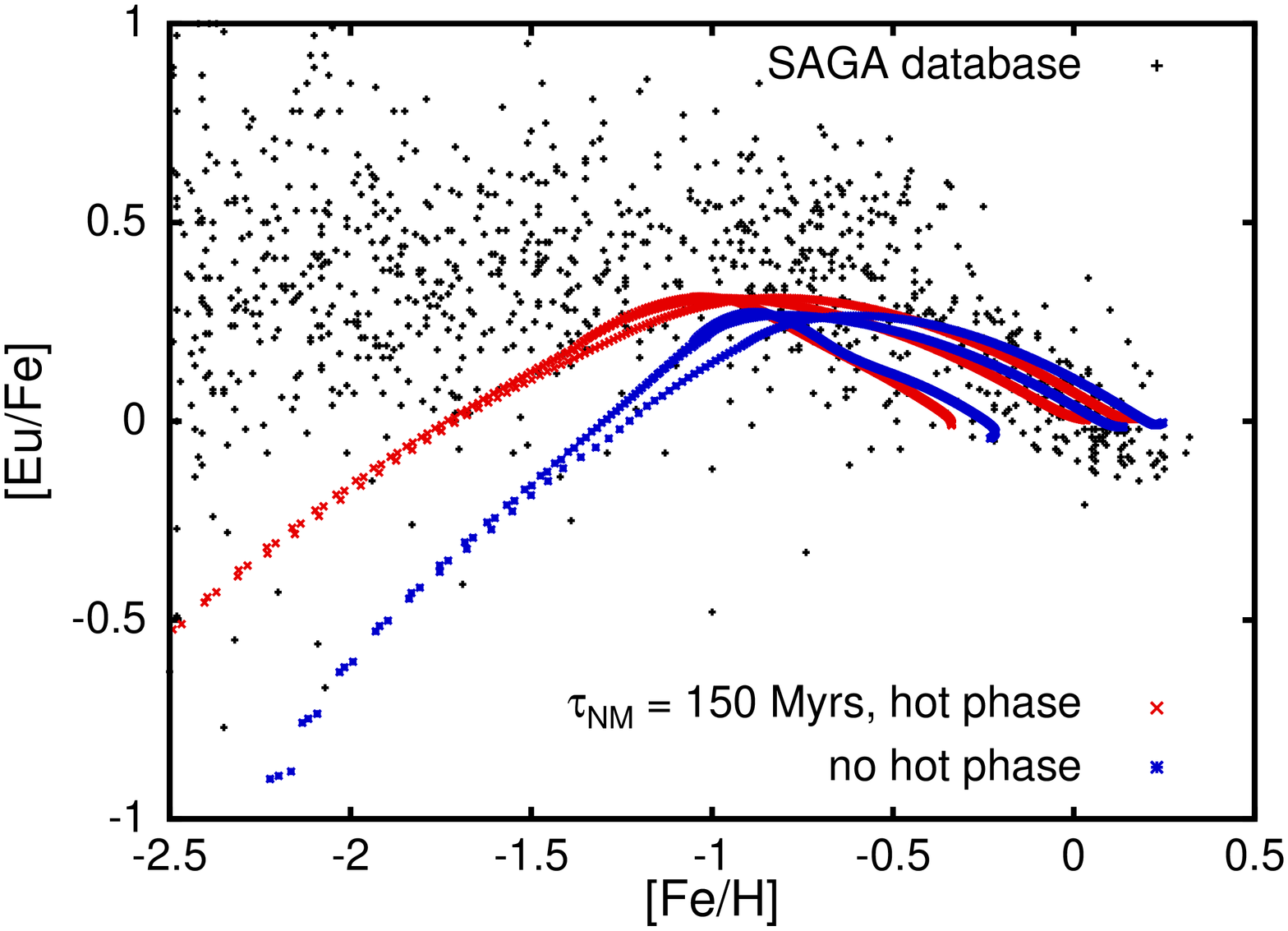,angle=-90,width=\hsize}
\epsfig{file=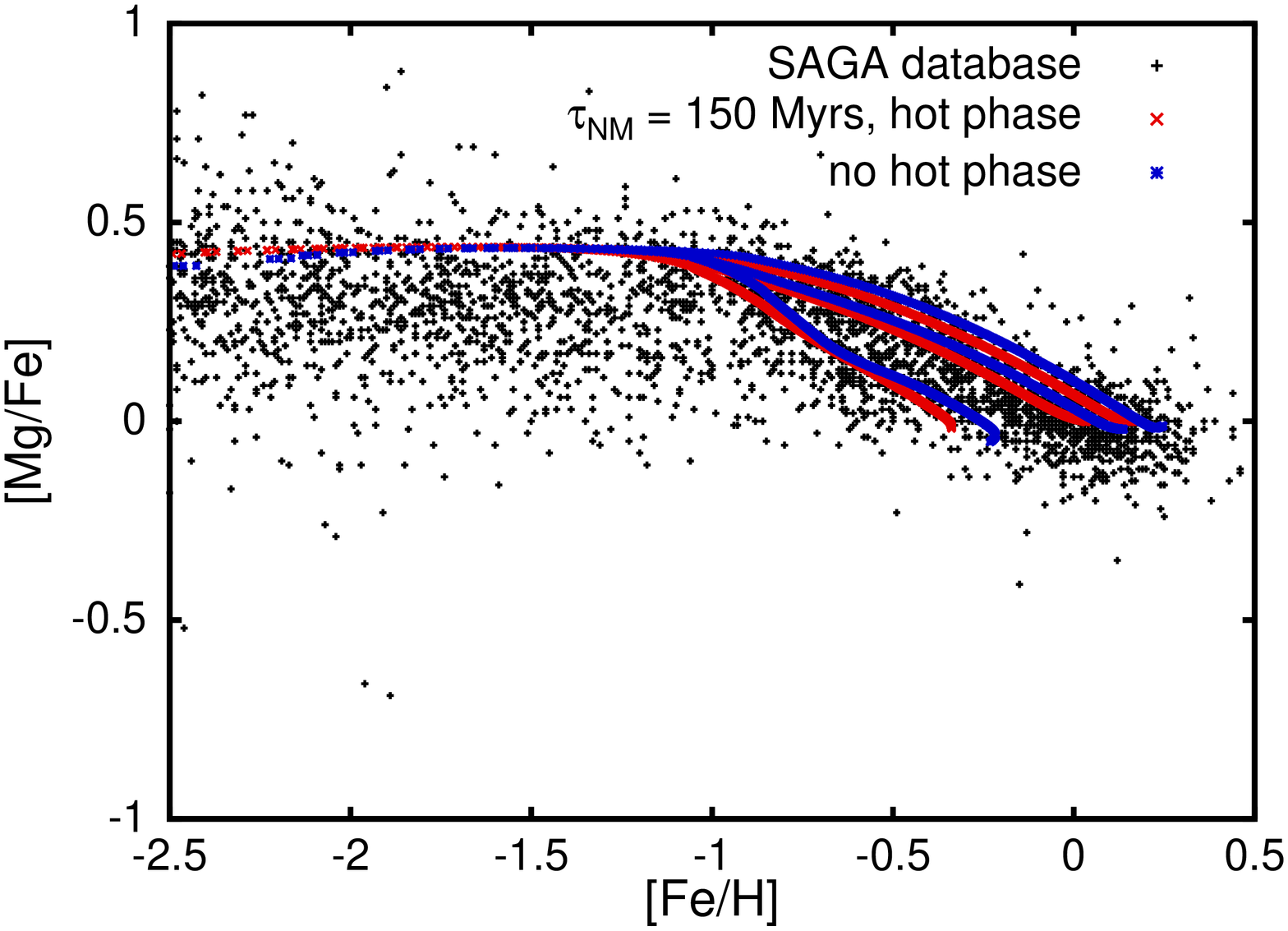,angle=-90,width=\hsize}
\caption{Introducing the 2-phase ISM: the top panel shows $\eufe$/r-process abundance vs. $\feh$ modelled with the same model and trajectories as in Fig.~\ref{fig:modelnohot}. The bottom panel shows again $\mgfe$ for comparison. In all trajectories, we use $\tau_\NM = 150 \Myr$. The blue points show the model with only one gas phase, as in the previous plot. The red points depict a model where we introduce the hot ISM through which stellar yields have to cycle, while permitting a fraction of $\fcNM = \fcccSN = 0.25$ to directly enter the cold star-forming ISM.}\label{fig:modelhotnohot}
\end{figure}

With Fig.~\ref{fig:modelhotnohot} we approach the theme of this paper. The top and bottom panels show again the evolution of $\eufe$ and $\mgfe$. In comparison to the previous results we now introduce (model shown with red points) the hot phase of the ISM, and assume that most yields are fed into the hot ISM from which they have to cool down back into the cold gas phase. Compared to the 1-phase model, all trajectories shift to the left, since lock-up in the hot phase significantly lowers the metallicity in the cold star-forming gas at any given time $t$ that is comparable to or shorter than the cooling time $\taucool$. In contrast to the $\eufe$-$\feh$ plane, the trajectories in the $\mgfe$-$\feh$ plane are only weakly affected, since the SNIa timescale, which delays the additional Fe yields, is larger than $\taucool$. This again confirms that our choice of a low value of $\fcSNIa$ has no practical consequence to this paper. For $\eu$ the situation is different: the neutron star merger timescale fulfills $\tauNM \ll \taucool$, so here the lock-up becomes important. This shifts the initial (roughly linear, see Appendix A1, eq.~\ref{eqn:early}) rise of the model trajectories in the $\eufe$-$\feh$ plane to the left, deeper into the regime dominated by stochastic chemical evolution. Further, along the thick disc plateau, there is also a very mild increase in $\eufe$. It derives from the change in star formation rates, the different effect of NM delay time with the hot gas phase, and a small upward nudge at late times due to the lower $\fcSNIa$.  However, this effect is still far too small to enable this prescription to break the hard limits discussed in Section~\ref{sec:analytical}.  

\begin{figure}
\epsfig{file=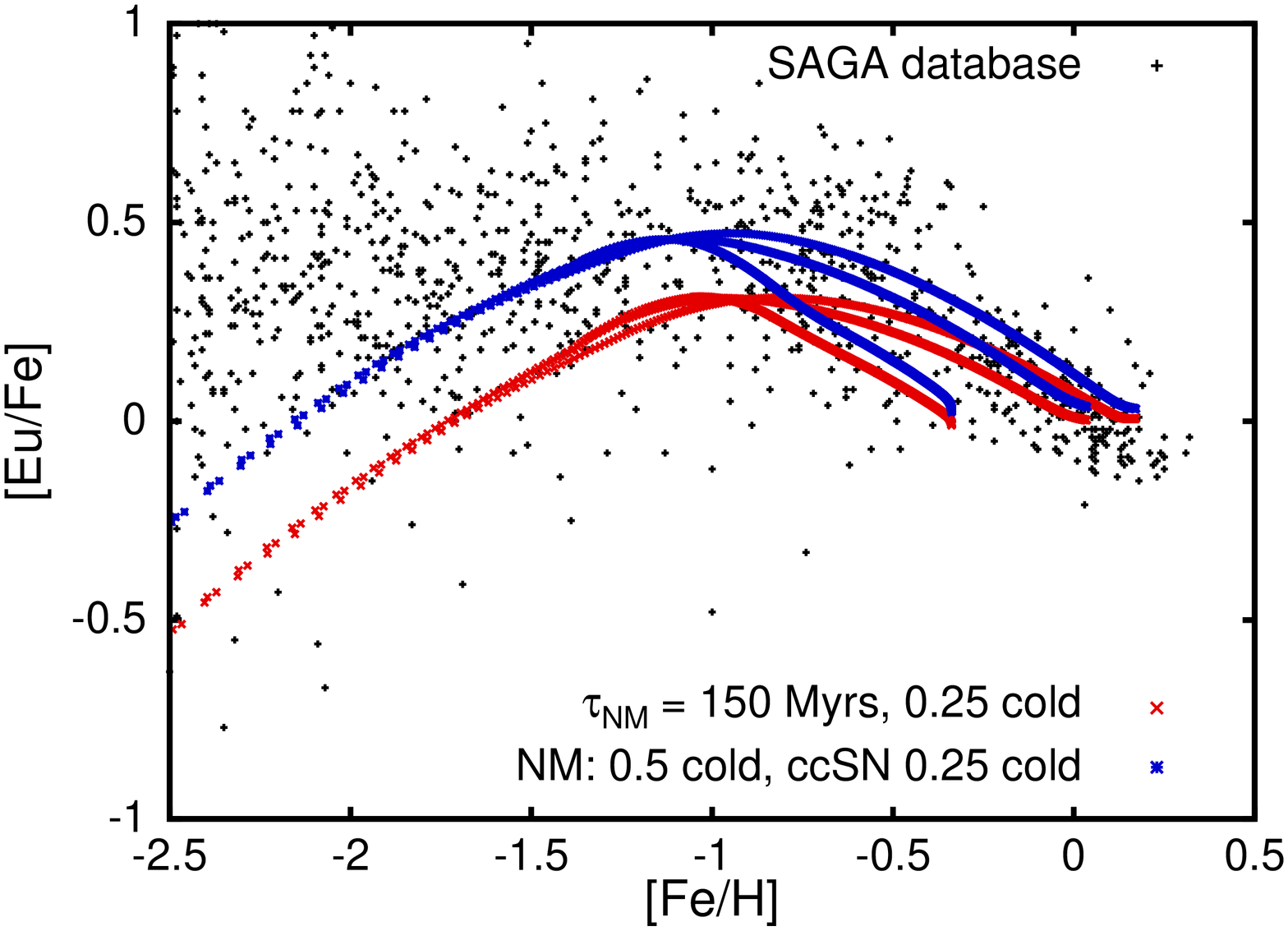,angle=-90,width=\hsize}
\epsfig{file=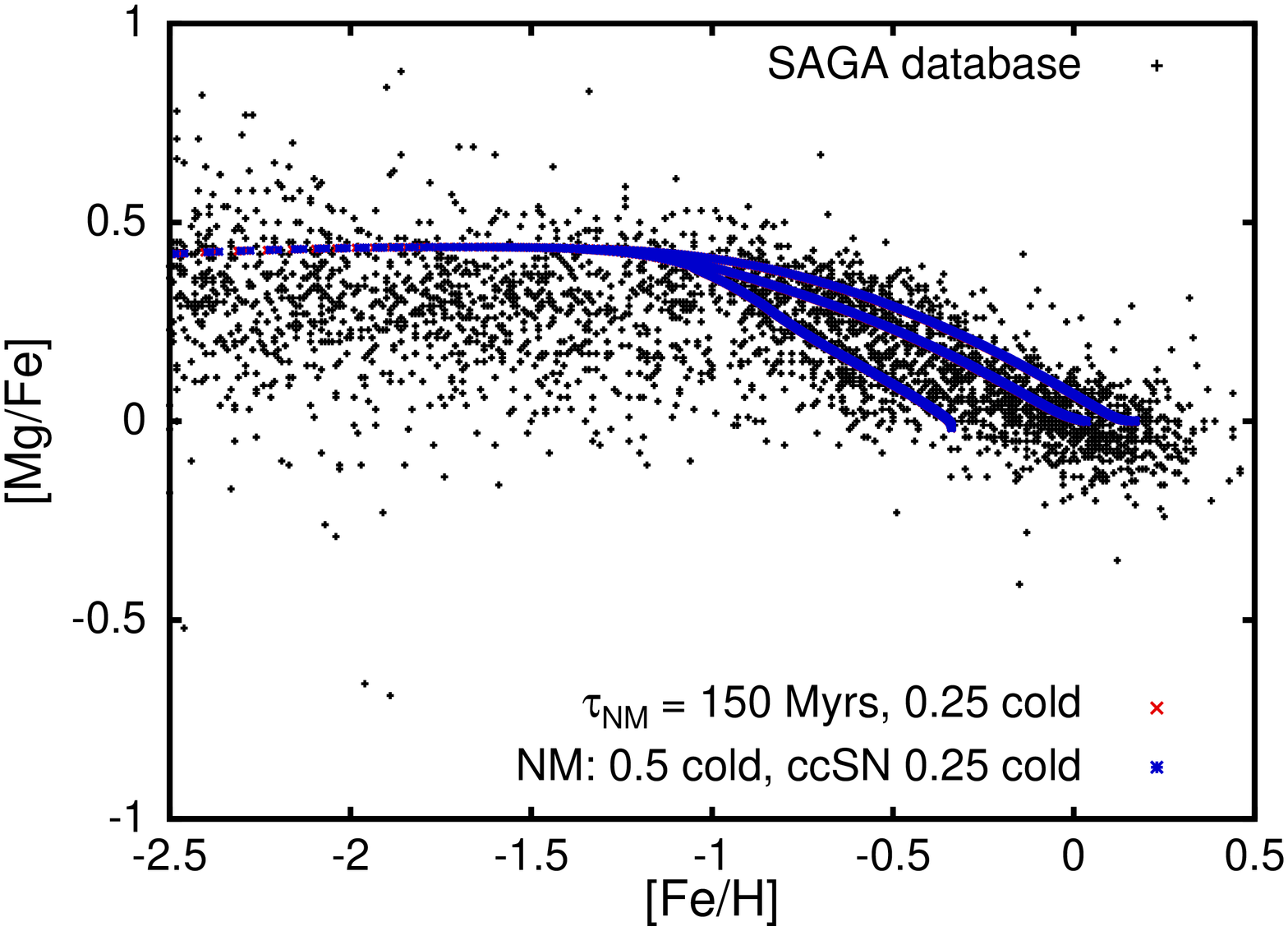,angle=-90,width=\hsize}
\caption{Testing differences in how different sources enter the ISM. We keep $\tau_\NM = 150 \Myr$ constant, keep all other parameters the same, and fix the fraction of ccSN yields entering the cold phase of the ISM directly to $\fcccSN = 0.25$. However, we now vary the fraction $\fcNM$ of neutron star merger yields entering the cold gas phase directly. As predicted with our simple analytical equations in eq. (\ref{eq:eufe2phase}), this permits a marked increase of the $\eufe$ abundance in the thick disc plateau.
}\label{fig:modelhotspec}
\end{figure}

The possible solution is shown in Fig.~\ref{fig:modelhotspec}. While keeping all other parameters constant in the 2-phase model presented in Fig.~\ref{fig:modelhotnohot}, we now allow a larger fraction of neutron star merger yields to enter the cold ISM than the yields from ccSNe, i.e.: $\fcNM > \fcccSN$. Just as predicted in the simple analytical picture of equation (\ref{eq:eufe2phase}), we now obtain plateau values fully in agreement with the $\eufe$ observations. 

We stress that the goal is not to match the observations at low metallicities: our models predict very few thick disc stars formed in the linearly rising section of the model trajectories at $\feh \lesssim -1.5 \dex$. The stellar populations with $\feh \lesssim -1.5 \dex$ are dominated by accreted objects, i.e. stars from dwarf galaxies. At first order, simply imagine the presented model to be shifted to the left, as those smaller galaxies have lower star formation efficiencies and larger loss rates to the IGM. Many of them will also cease to form stars before evolving to low $\mgfe$ and $\eufe$ values. We thus can expect that the high $\eufe$ plateau is extended to $\feh \lesssim -1.5 \dex$ by a superposition of analogous dwarf galaxy models. Fitting this region would also not add new information to our case: the low-$\feh$ regime is not only composed of a large number of different systems, but the different flavours of stochastical chemical evolution, and uncertainties e.g. about the incidence of magneto-rotational ccSNe, add a large space of additional parameters that do not pertain to our question, which is answered on the more tightly constrained thick disc stars.

\begin{figure*}
\epsfig{file=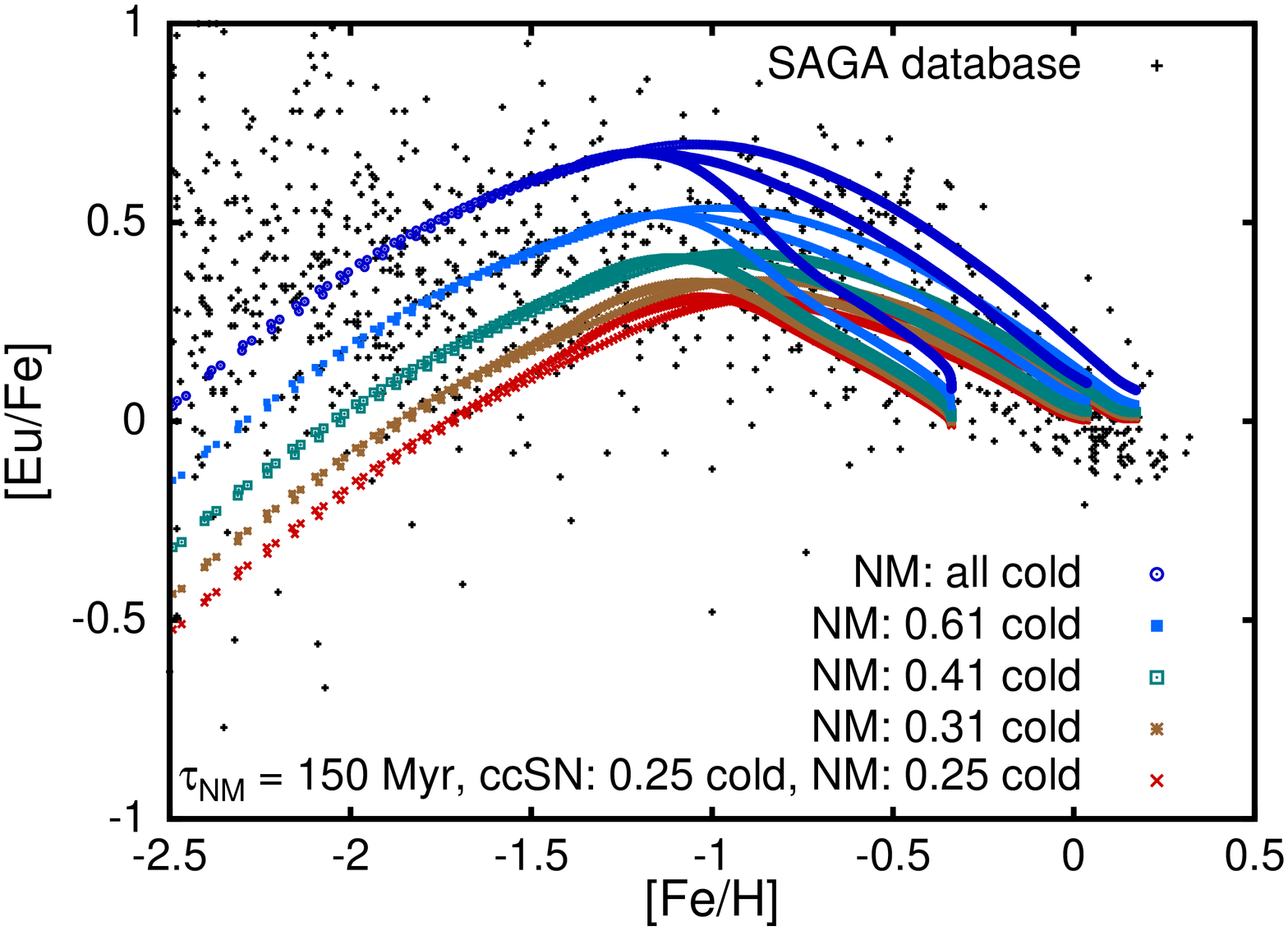,angle=-90,width=0.49\hsize}
\epsfig{file=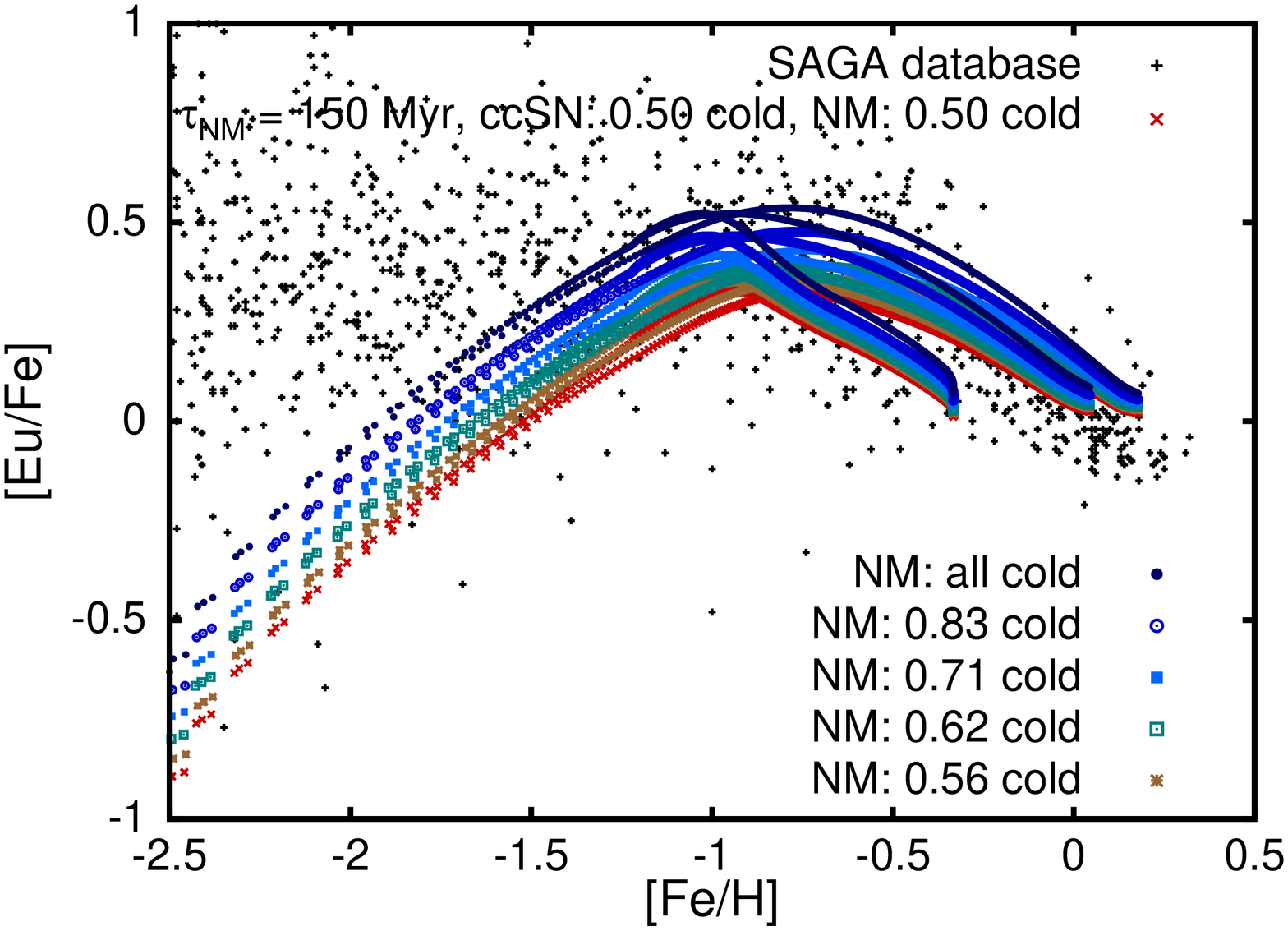,angle=-90,width=0.49\hsize}
\epsfig{file=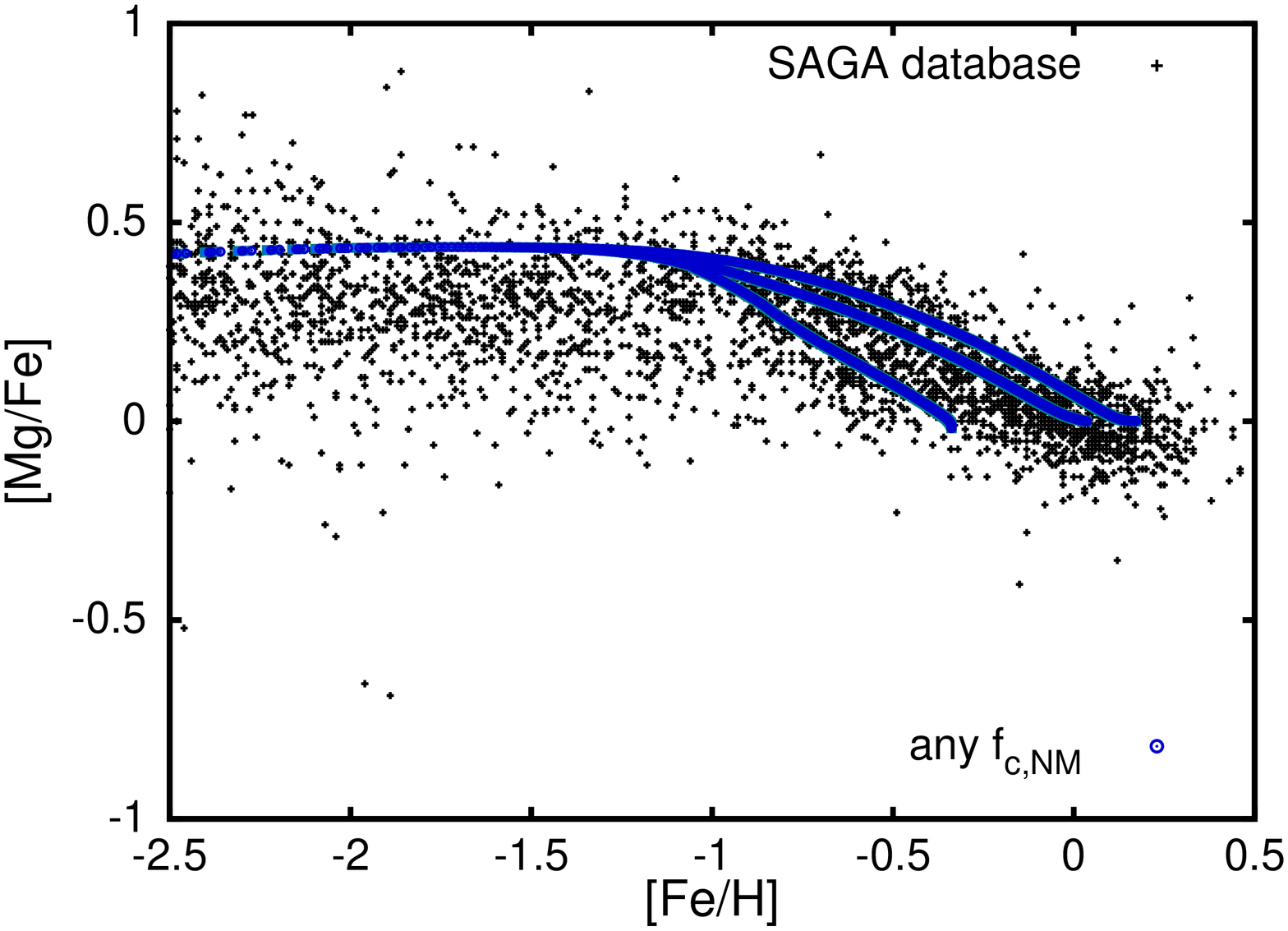,angle=-90,width=0.49\hsize}
\epsfig{file=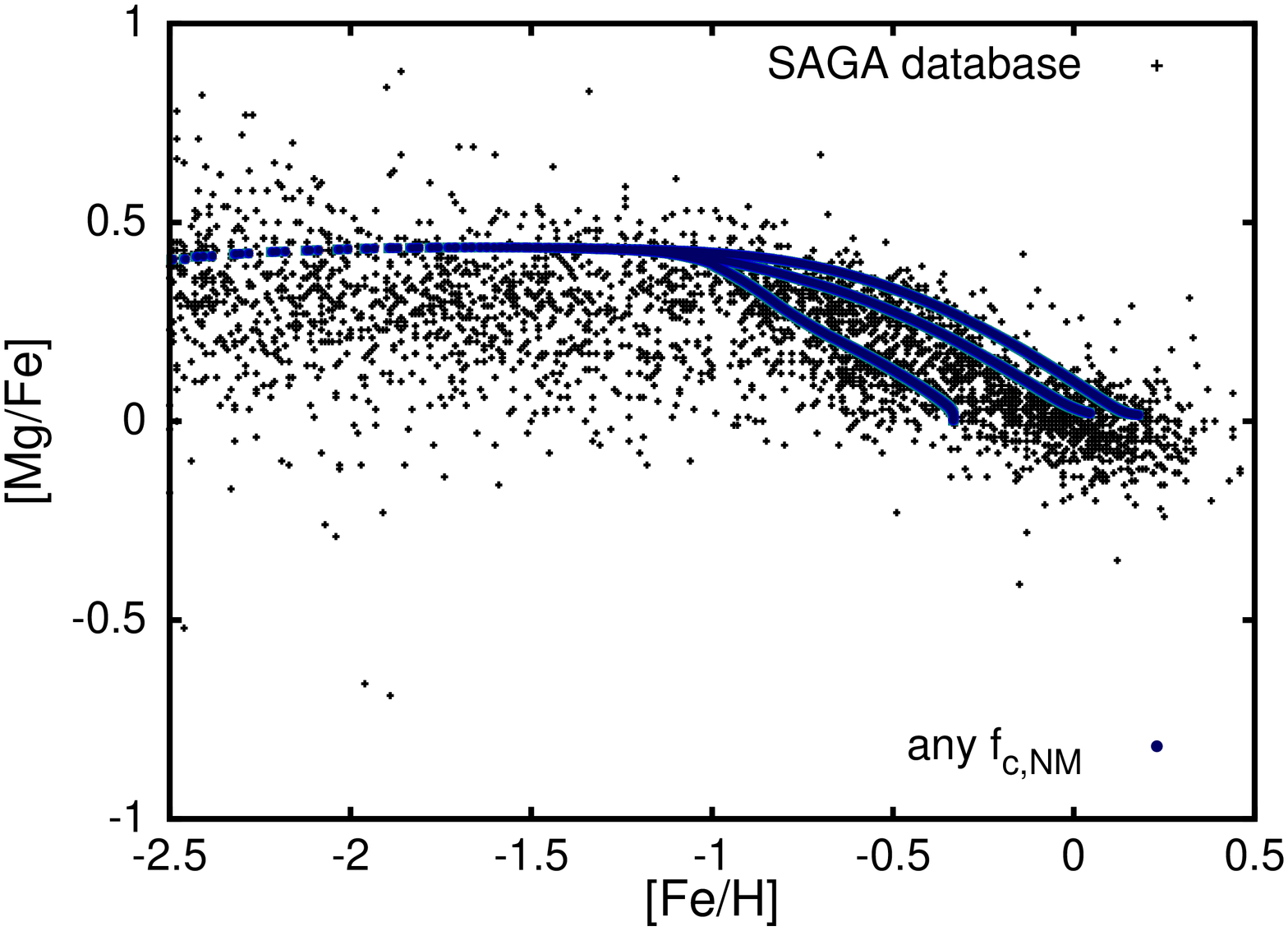,angle=-90,width=0.49\hsize}
\caption{In an extension of Fig.~\ref{fig:modelhotspec}, we provide different values of $\fcNM$ for a fixed $\fcccSN = 0.25$ (left-hand column) and $\fcccSN = 0.5$ (right column).}\label{fig:modelspecvar}
\end{figure*}

In Fig.~\ref{fig:modelspecvar} we present an extension of Fig.~\ref{fig:modelhotspec}. We now present for $\eufe$ (top row) and $\mgfe$ (bottom row) a range of values for $\fcNM$ for two possible values of $\fcccSN = 0.25$ (left-hand column) and $\fcccSN = 0.5$ (right-hand column). The right-hand side should be considered only as an instructive comparison, as we do not favour such a large cold ISM channel from ccSNe. We can see by the comparisons that eq. (\ref{eq:eufe2phase}) correctly predicts the behaviour. Multiplying $\fcccSN$ by a factor $2$ shifts $\eufe_{\rm pl}$ downwards by $0.3 \dex$. Similarly at fixed $\fcNM$ the plateau would shift upwards by $0.3 \dex$ compared to the left-hand column of Fig.~\ref{fig:modelspecvar} if we chose $\fcccSN = 0.125$. The same shift applies for every factor $2$ in $\fcNM$. A slight caution needs to be taken as we can see from the high-metallicity/low-$\eufe$ end of the trajectories. For large values $\fcNM$, less r-process elements stay locked in the hot phase, which would require a mild correction to our adopted normalisation to bring the trajectories back to $\eufe = 0$. However, this is a small fraction of the change in the plateau, and readers can easily convince themselves that the difference between the trajectory-endpoints and the plateau rises strongly with $\fcNM$ irrespective of the correction.

\begin{figure}
\epsfig{file=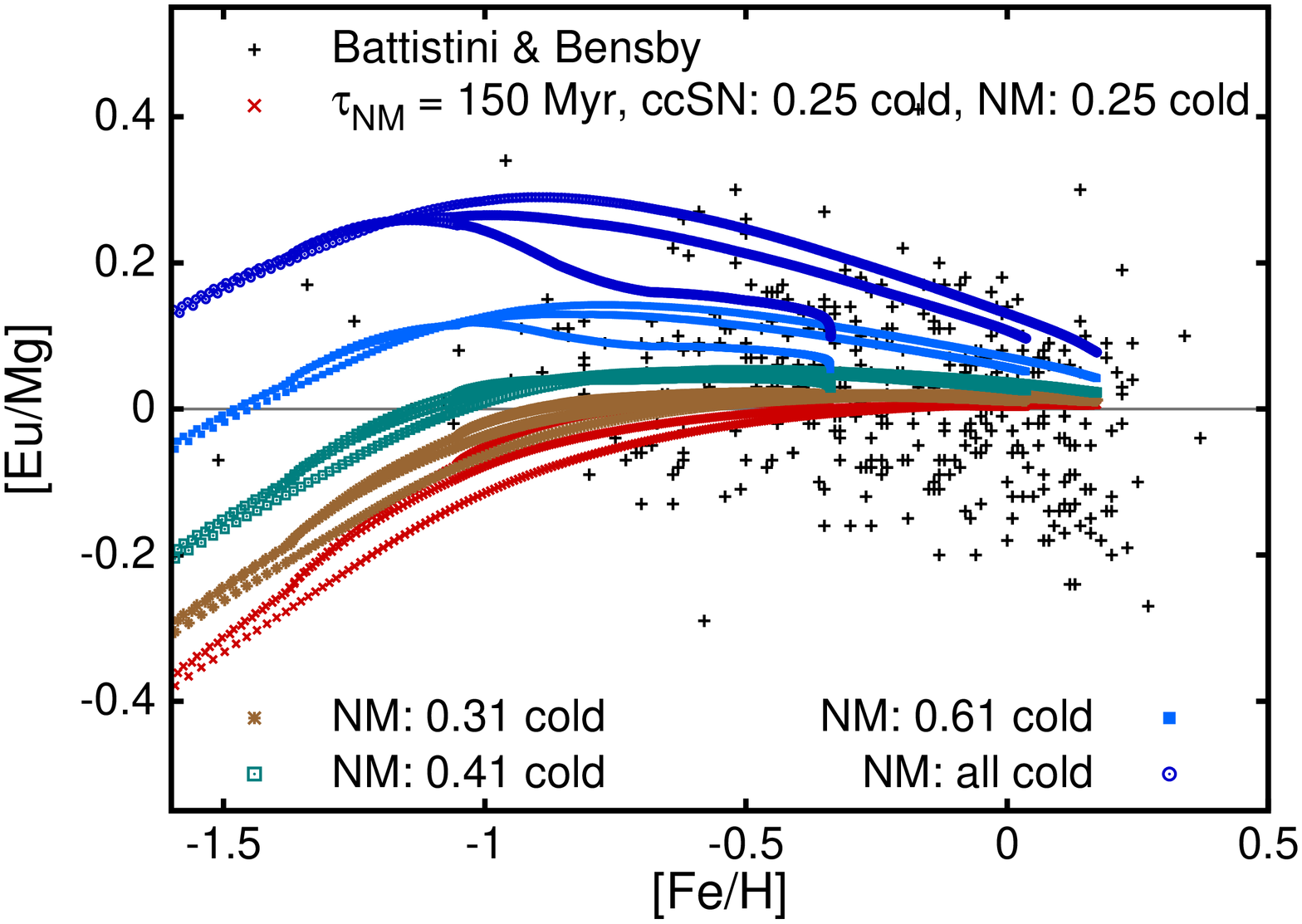,angle=-90,width=\hsize}
\epsfig{file=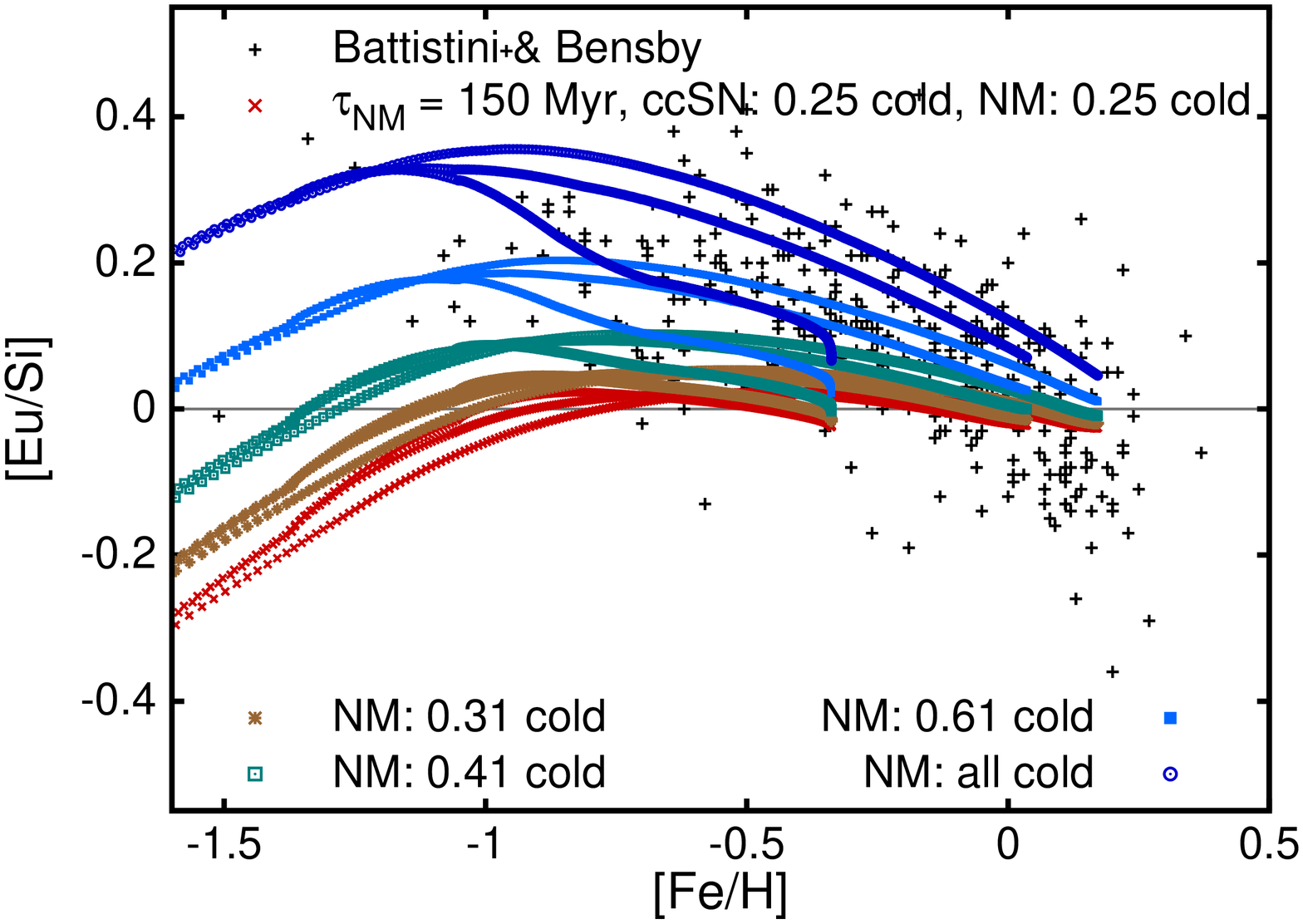,angle=-90,width=\hsize}
\caption{For the same models as in Fig.~\ref{fig:modelhotspec}, we provide the trajectories in $\eumg$ (top panel) and $\eusi$ (bottom) compared to the Battistini/Bensby data. Note that 1D-LTE analysis is implicated in over-estimating $\mg$ abundances (and thus under-estimating $\eumg$) by about $0.1 \dex$, which is also the amount by which the best-fitting model in $\eusi$ overestimates $\eumg$.}\label{fig:modelMgSispecvar}
\end{figure}

In analogy to Fig.~\ref{fig:modelMgSinohot}, Fig.~\ref{fig:modelMgSispecvar} analyses model trajectories in the $\eualpha$-$\feh$ plane when we vary $\fcNM$, affirming the points made above. When we allow for a larger fraction of neutron star merger yields to be contributed directly to the cold gas phase, the model values of $\eumg$ and $\eusi$ finally push significantly into positive territory. $\si$ requires a somewhat larger effect than $\mg$, which we again ascribe to the problem of $\mgh$ values being slightly overestimated in a 1D-LTE spectroscopic analysis. The figure also underlines again that different yields to the hot/cold gas phases readily account for the observed trends.

\begin{figure}
\epsfig{file=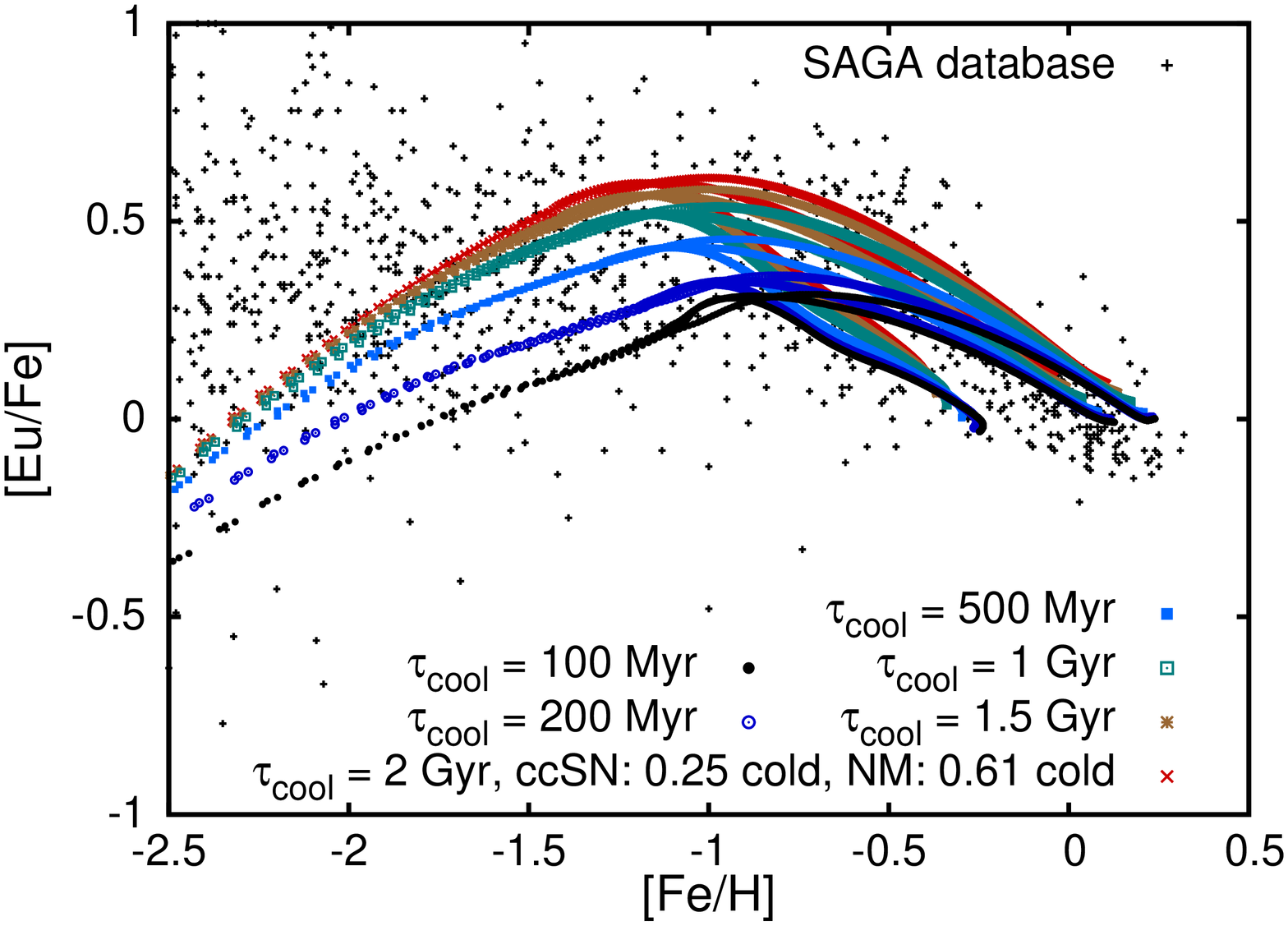,angle=-90,width=\hsize}
\epsfig{file=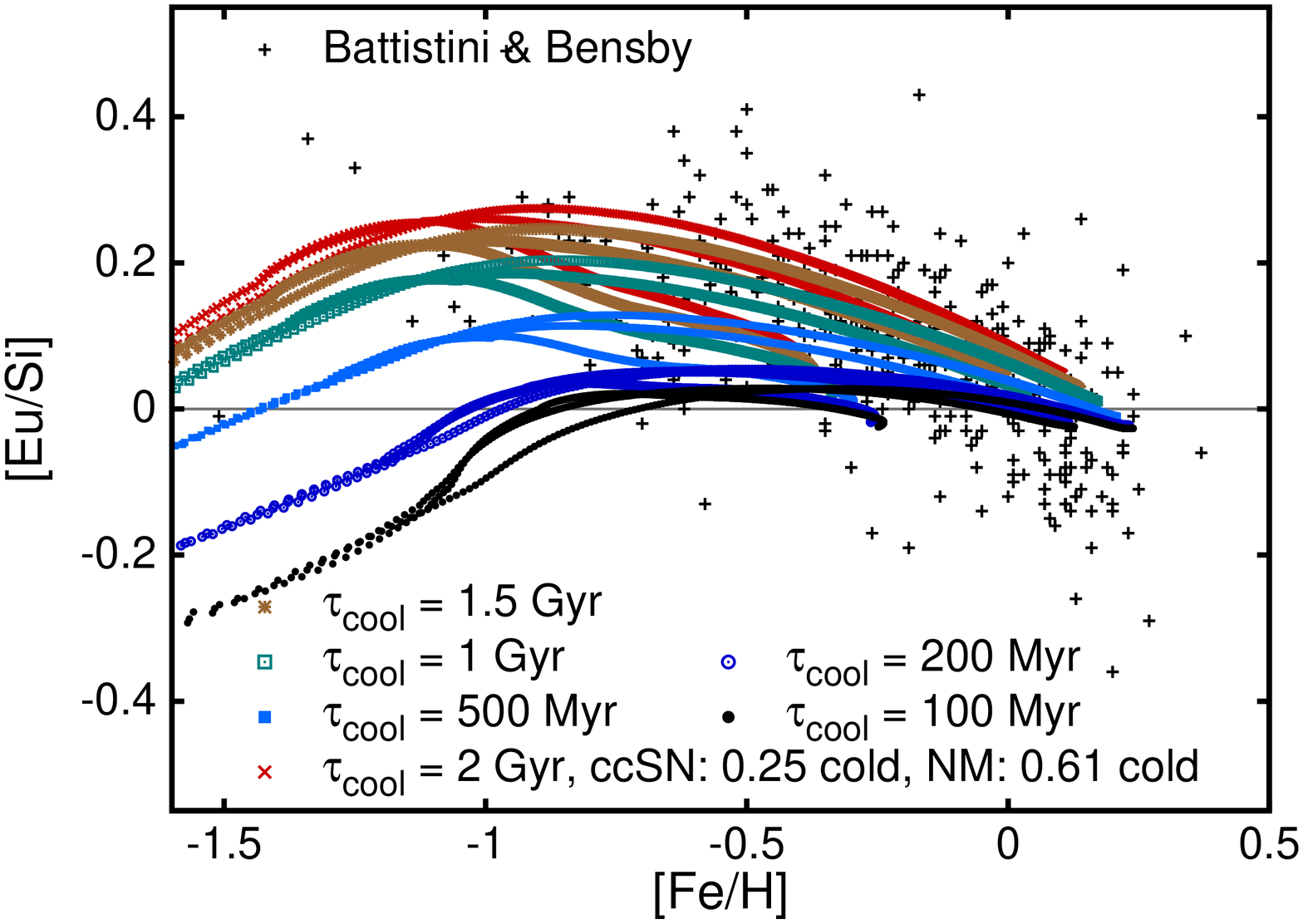,angle=-90,width=\hsize}
\caption{Testing different hot ISM cooling times for the model $\tau_\NM = 150 \Myr$, $\fcccSN=0.25$ and $\fcNM = 0.61$ for $\eufe$ vs. $\feh$ against the SAGA database (top panel) and $\eusi$ vs. $\feh$ against the Battistini \& Bensby sample (bottom panel). Note that we use different ranges on the x-axis.}\label{fig:modeltcool}
\end{figure}

Fig.~\ref{fig:modeltcool} tests different cooling times, $\tau_{\rm cool}$, for the hot ISM, for an otherwise standard model setting with $\fcccSN = 0.25$ and $\fcNM = 0.61$, showing the comparison with both the SAGA database in $\eufe$ (top) and the Battistini \& Bensby sample in $\eusi$ (bottom). We consider timescales below $\sim 500 \Myr$ and significantly above $2 \Gyr$ unlikely, since they would either approach values shorter than the orbital timescale or predict too much mass in the warm/hot ISM, but it is instructive to compare the model values. For very short cooling timescales the model of course reverts back to the behaviour of a 1-phase model as shown in Fig.~\ref{fig:modelnohot}. For very long timescales the model approaches the upper limit given in equation~\ref{eq:eufe2phase}, and the changes nearly saturate. We are not re-calibrating the model's loss rates and r-process yields between the different parameters -- one can see that the larger lock-up in models with large $\tau_{\rm cool}$ results in slightly lower end-point $\feh$ and larger $\eufe$. If we re-gauged the parameters, models with $\tau_{\rm cool} > 1 \Gyr$ would look nearly identical, though requiring slightly smaller yield losses and smaller r-process yield for larger $\tau_{\rm cool}$. When comparing to the SAGA database (top panel) we plot lower $\feh$ values. It is clear that $\tau_{\rm cool}$ is less important for the very low $\feh$ regime, where $\tau_{\rm cool}$ is large compared to the age of the stellar population. Most importantly, our assumption on $\tau_{\rm cool}$ impacts the estimated ratio of $\fcNM/\fcSNIa$, which is insensitive against changes in the region $\tau_{\rm cool} > 1 \Gyr$, but would have to be increased for shorter $\tau_{\rm cool}$, e.g. to about $4$ if one assumed $\tau_{\rm cool} = 500 \Myr$.

\begin{figure}
\epsfig{file=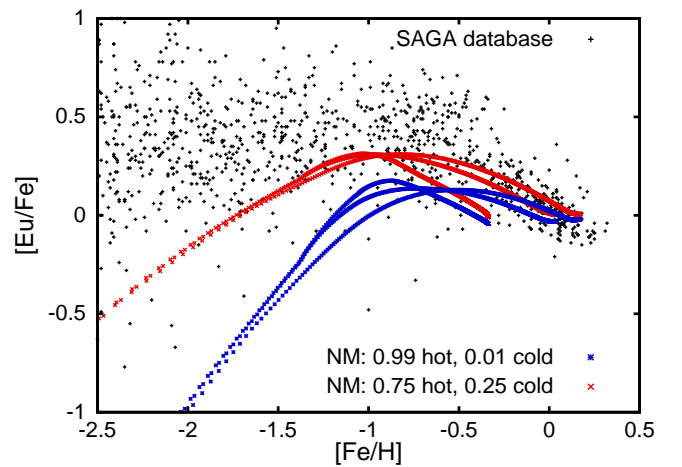,angle=-90,width=\hsize}
\caption{The model with a hot gas phase and $\fcNM = \fcccSN = 0.25$ compared to a model where we have no channel to the cold phase for r-process elements.}\label{fig:modelnogd}
\end{figure}

Readers could now argue that we do not know what the relative contributions to the respective gas phases of each yield class are, and they would be right. The point is that we can a priori expect differences, i.e. $\fcNM \neq \fcccSN$, but we have no way to quantify them at present, apart from looking at the abundance plane. Thus, to complement this discussion, Fig.~\ref{fig:modelnogd} compares the chemical evolution in a model with both gas phases and treating the neutron-star mergers like ccSNe, i.e. $\fcNM = \fcccSN = 0.25$ to a model with both gas phases in which neutron-star merger yields go entirely into the hot phase. Obviously the latter model strongly exacerbates the problem that neutron-star mergers contribute their yields later than the ccSNe, and the low $\eufe$ trajectories strongly disfavour this scenario, even if we could find other ways (e.g. metallicity dependent yields) to help raise the $\eufe$ ratios at early times.

\section{Conclusions}\label{sec:Conclusions} 

We should take three main points from this analysis: 
\begin{itemize}
\item Modelling the different phases of the interstellar medium (hot vs. cold star forming gas) is vital to understand chemical evolution on timescales smaller than $\sim 1 \Gyr$.
\item In contrast to 1-phase chemical evolution models, we find that neutron star mergers as source of r-process elements with reasonable delay time distributions (delay times of order $100 \Myr$) can explain the observed abundance patterns, provided that the fraction of NM yields delivered directly to the cold star-forming phase of the ISM is higher than that of ccSN yields.
\item The only other significant sources of r-process elements are (possibly) ccSNe, but we can only explain the $\eusi > 0$ and $\eumg > 0$ values in the thick disc if there is a source of r-process elements that differs from a constant ccSN contribution. This implies a significant source of r-process elements besides ccSNe - most naturally neutron star mergers.
\end{itemize}

The main problem is not as much explaining r-process abundances at the very metal-poor end, where stochastic chemical evolution and the superposition of stellar populations from different accreted dwarf galaxies complicate the picture (and provide freedom in parameter choice). Instead, the challenge is the relatively high-metallicity edge of the high $\eufe$ sequence, near the \enquote{knee} of the $\mgfe$ vs. $\feh$ distribution. 

A cautionary point that would allow us to lower the NM contribution to the r-process budget is the possibility of metallicity dependent yields from ccSNe/collapsars. We did not have space here for an in-depth discussion. However, these models require a lot more fine-tuning and face three observational challenges that are not yet convincingly resolved: a significant increase of r-process yields at low metallicities would have to i) still conform with normal $\eufe$ values observed in dwarf galaxies, and to ii) be fine-tuned in order not to incline the thick disc plateau. Further, a collapsar model with a metallicity dependence on the metal-rich end, as suggested in \cite{Siegel18} would have to still produce the radial $\eufe$ profile measured in the Milky Way. 

We have shown with simple analytical considerations in Section~\ref{sec:analytical}, a more comprehensive analytical model in Appendix A1, and with the chemodynamical models of \cite{SM17} in Section~\ref{sec:specmodel}, that $\eualpha > 0$ or $\eufe > \afe$, cannot be achieved in a 1-phase chemical evolution model in general. However, they are readily reproduced with a simple chemical evolution model as soon as we account for a 2-phase ISM and allow yields from neutron star mergers to enter the cold, star forming gas at a mildly higher fraction than the ccSN yields, $\fcNM > \fcccSN$. The factor $\fcNM / \fcccSN$ required here can be directly estimated from the $\eualpha$ value of the thick disc knee to be about $0.2 \dex$ or $\fcNM / \fcccSN \approx 2$.

We find that the this factor is quite insensitive to cooling timescales $\tau_{\rm cool} \gtrsim 1 \Gyr$ and would need to be increased for shorter $\tau_{\rm cool}$. There is hope to derive better constraints for $\tau_{\rm cool}$ from the chemical evolution of elements produced on different timescales and by different sources, e.g. s-process elements.

The slightly different behaviour of $\sife$ and $\mgfe$ in observations serves as a reminder that one should not blindly trust observed abundance trends. In particular, trends vs. $\feh$ are vulnerable to increasing deviations from the frequently assumed LTE, when with lower metallicity the importance of collisions in the photosphere's plasma decreases. Similarly, we calibrated this model on $\mgfe$ and thus have a mismatch by $\sim ~ 0.07 \dex$ with $\sife$. This has no impact on our qualitative results, but it puts systematic uncertainty on the exact value needed for $\fcNM/\fcccSN$. Further, the qualitatively consistent behaviour of $\eualpha > 0$ and $\eufe > \afe$ for several different $\alpha$ elements makes it unlikely that the problem is a mere result of our still incomplete understanding of stellar spectra. Non-LTE analyses of $\eu$ are very rare, though from \cite{Zhao16} it seems that $\eu$ and $\eufe$ measurements are quite robust and if at all increase vs. LTE analysis. More systematic studies are quite sorely needed. 

The re-distribution of r-process elements between different gas phases and the resulting changes to abundance trends might at first hand seem like an annoying complication, but it may be turned into an important/useful diagnostic if simulations have an appropriate recipe for hot/cold gas dynamics and feedback. In this light, the strong discrepancy in papers like \cite{Voort15} between the modelled and observed trends of $\eufe$ vs. $\feh$ in the abundance plane is a chance to improve on the ISM modelling in hydrodynamical simulations, and our paper may provide an analytic framework to understand the differences between these and other hydro simulations that matched the data better \citep[e.g.][]{Shen15}. 

We point out that the 2-phase ISM assumed here is a simplified picture of a real galaxy, which will have a whole range of delay times that will vary with how far from the star-forming regions material is expelled and how long it takes for yields to re-enter the star-forming phase. So some of the difference in $\fcNM$ vs. $\fcccSN$ may be due to ccSN yields being expelled further from the disc.

More generally, we conclude that observed r-process abundance trends provide empirical evidence for both an important NM contribution to r-process enrichment and differential distribution of yields to the cold and hot phases of the ISM.

\section*{Acknowledgements}

We thank our referee F. Thielemann for enlightening comments. It is a pleasure to thank Eliot Quataert for pointing out the problem of neutron star chemical evolution. We would like to thank M. Bergemann, J. Binney, J. Johnson, J. Magorrian, B. Metzger, N. Prantzos, D. Siegel, and S. Smartt for helpful comments. RS is supported by a Royal Society University Research Fellowship. This work was performed using the Cambridge Service for Data Driven Discovery (CSD3), part of which is operated by the University of Cambridge Research Computing on behalf of the STFC DiRAC HPC Facility (www.dirac.ac.uk). The DiRAC component of CSD3 was funded by BEIS capital funding via STFC capital grants ST/P002307/1 and ST/R002452/1 and STFC operations grant ST/R00689X/1. DiRAC is part of the National e-Infrastructure.

\bibliographystyle{mnras}
\bibliography{paper}

\begin{figure*}
\epsfig{file=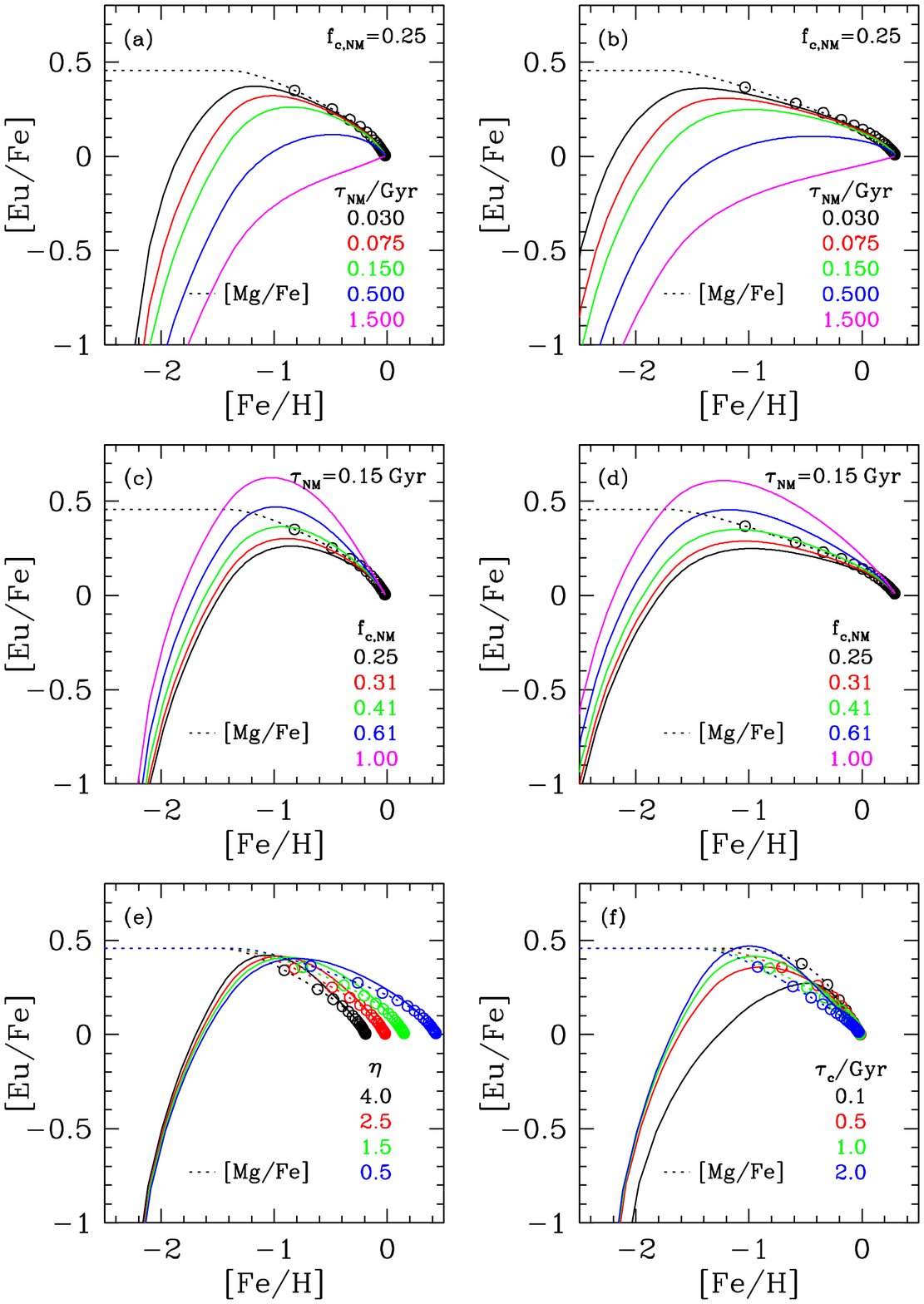,angle=0,width=0.9\hsize}
\caption{Evolution of [Eu/Fe] (solid lines) and [Mg/Fe] (dotted lines, with circles
plotted at 0.5 Gyr intervals)) vs. [Fe/H],
in one-zone analytic models following equations (9) and (10).
(a) Models with $\eta=2.5$, $\fcNM=\fcccSN=0.25$, $\tau_c=1.0\,{\rm Gyr}$
and varying $\tau_{\rm NM}$.
(b) Same as (a) but with $\eta=0$ and all yields multiplied by 0.4.
(c) Models with $\tau_{\rm NM}=0.15\,{\rm Gyr}$ and varying $\fcNM$.
(d) Same as (c) but with $\eta=0$ and all yields multiplied by 0.4.
(e) Models with $\fcNM=0.5$, $\fcccSN=0.25$, and varying $\eta$.
(f) Same as (e), but with $\eta=2.5$ and varying $\tau_c$.}\label{fig:analytic}
\end{figure*}

\begin{figure*}
\epsfig{file=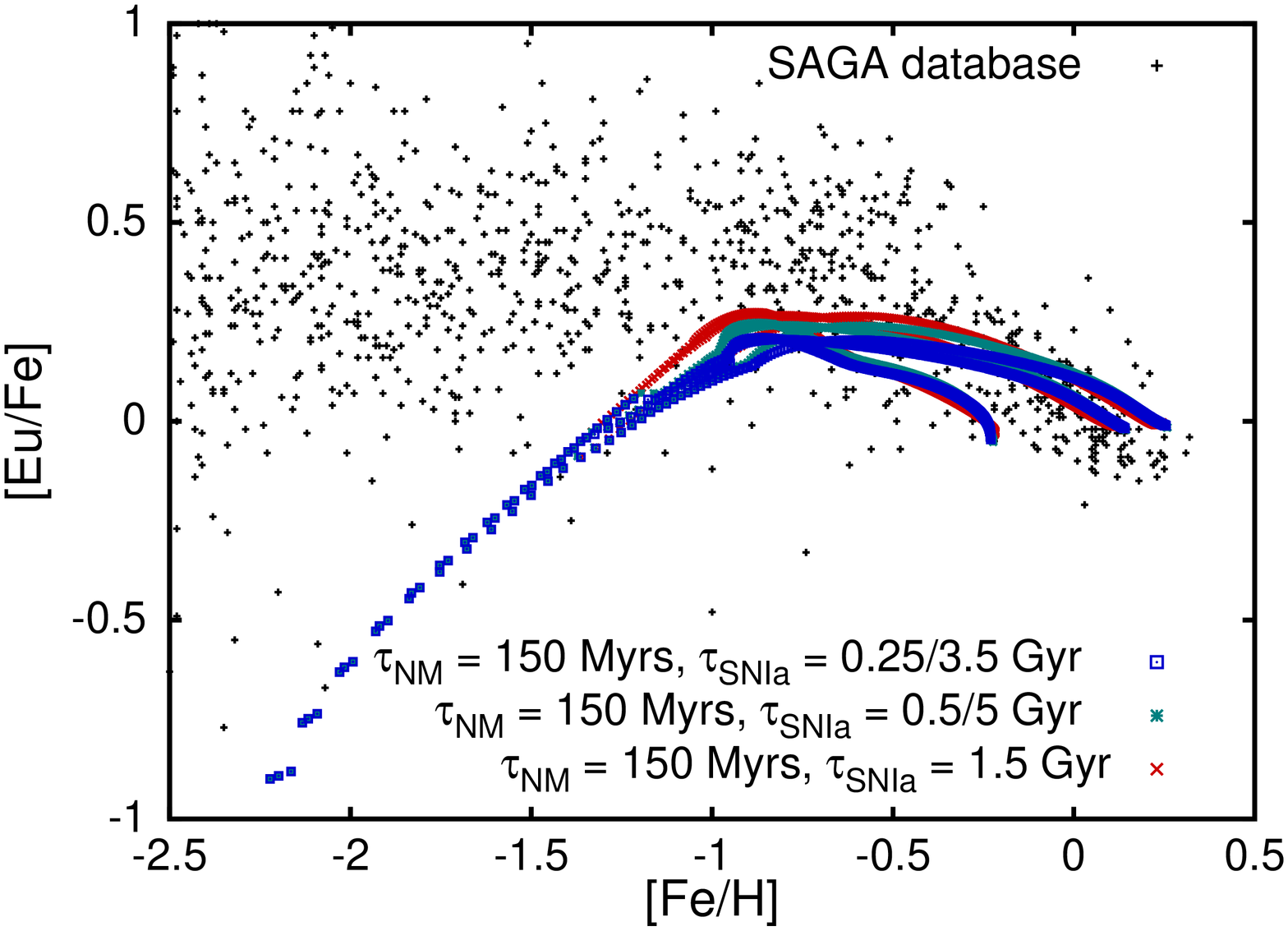,angle=-90,width=0.49\hsize}
\epsfig{file=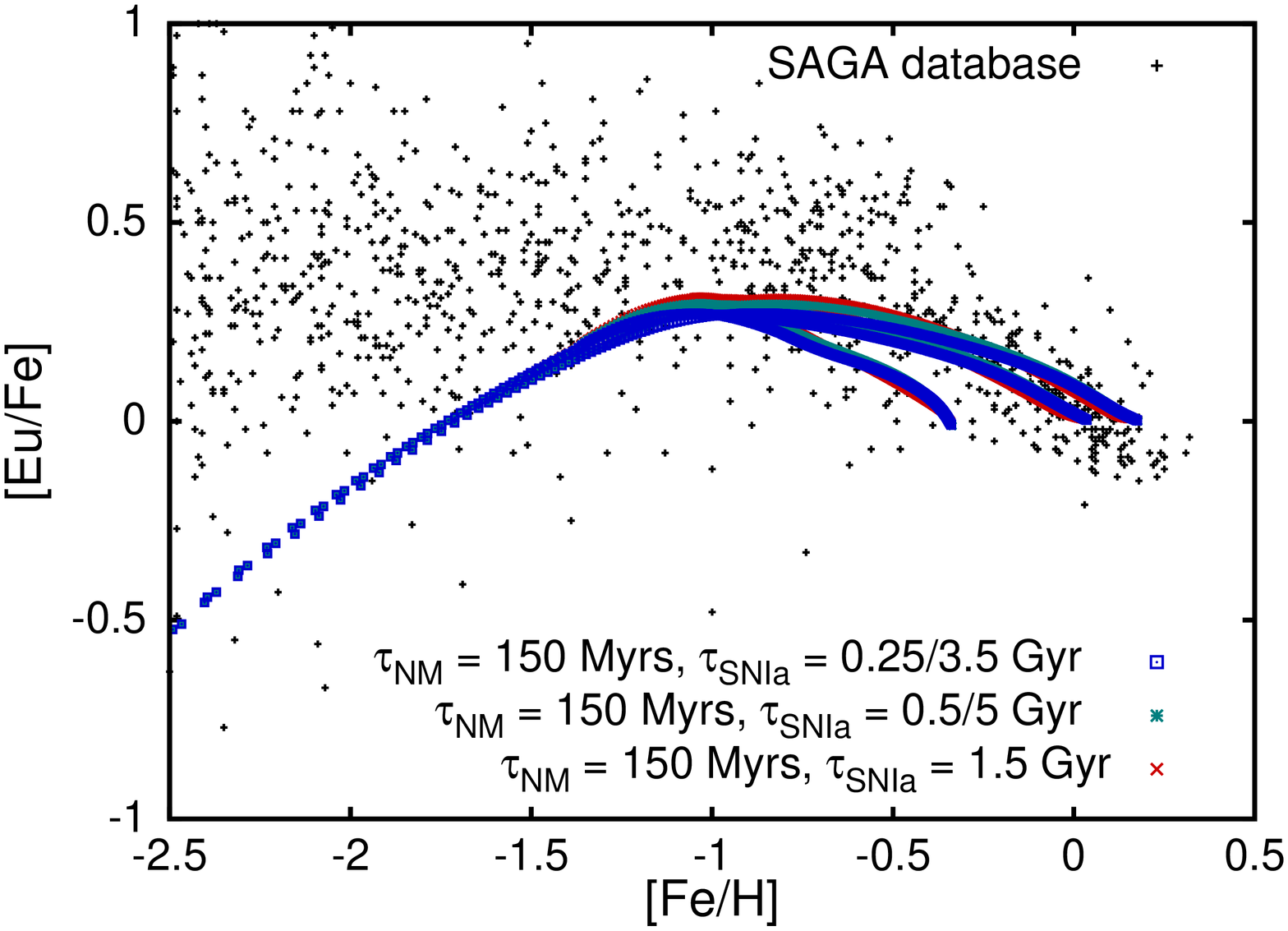,angle=-90,width=0.49\hsize}
\epsfig{file=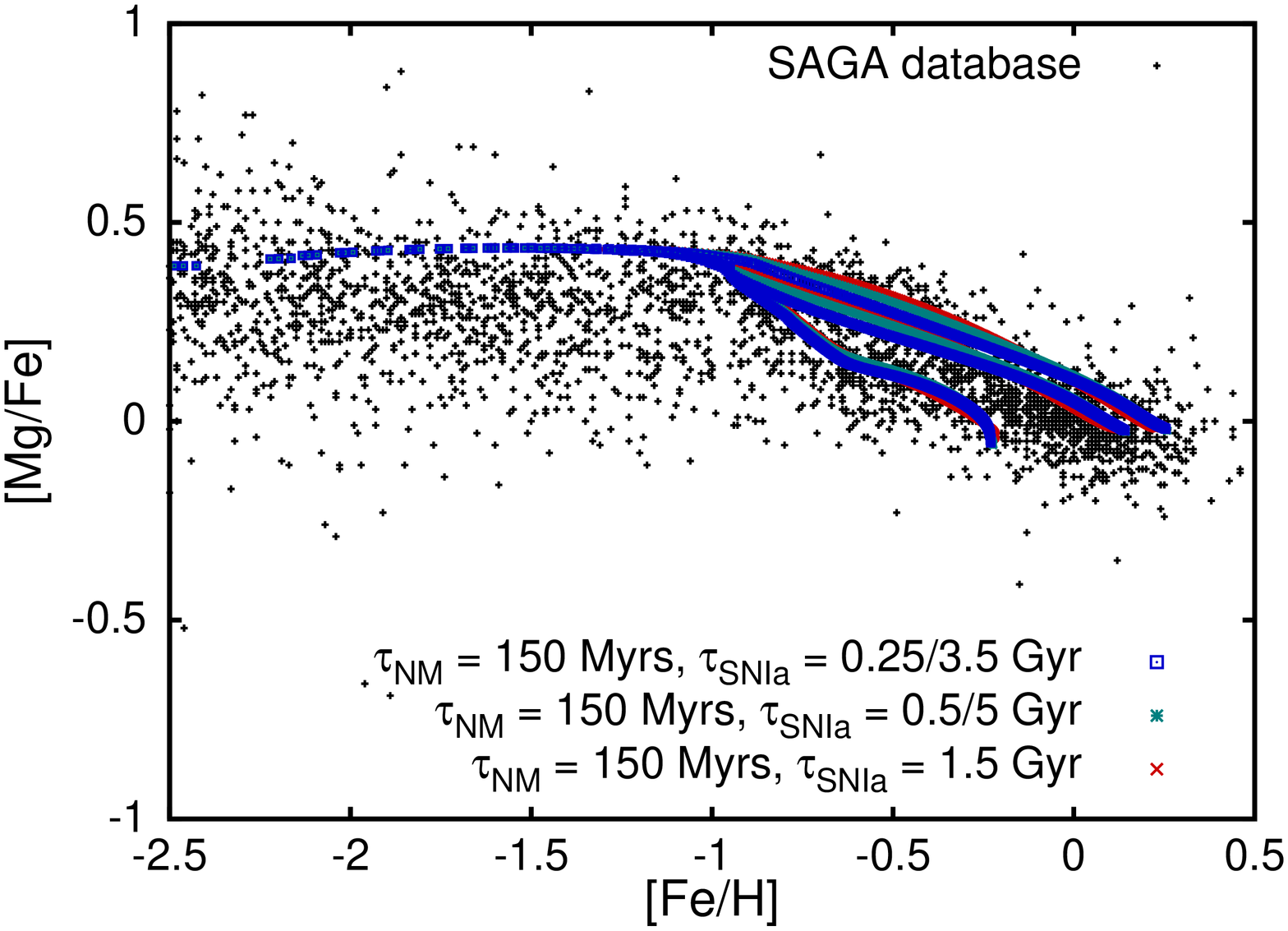,angle=-90,width=0.49\hsize}
\epsfig{file=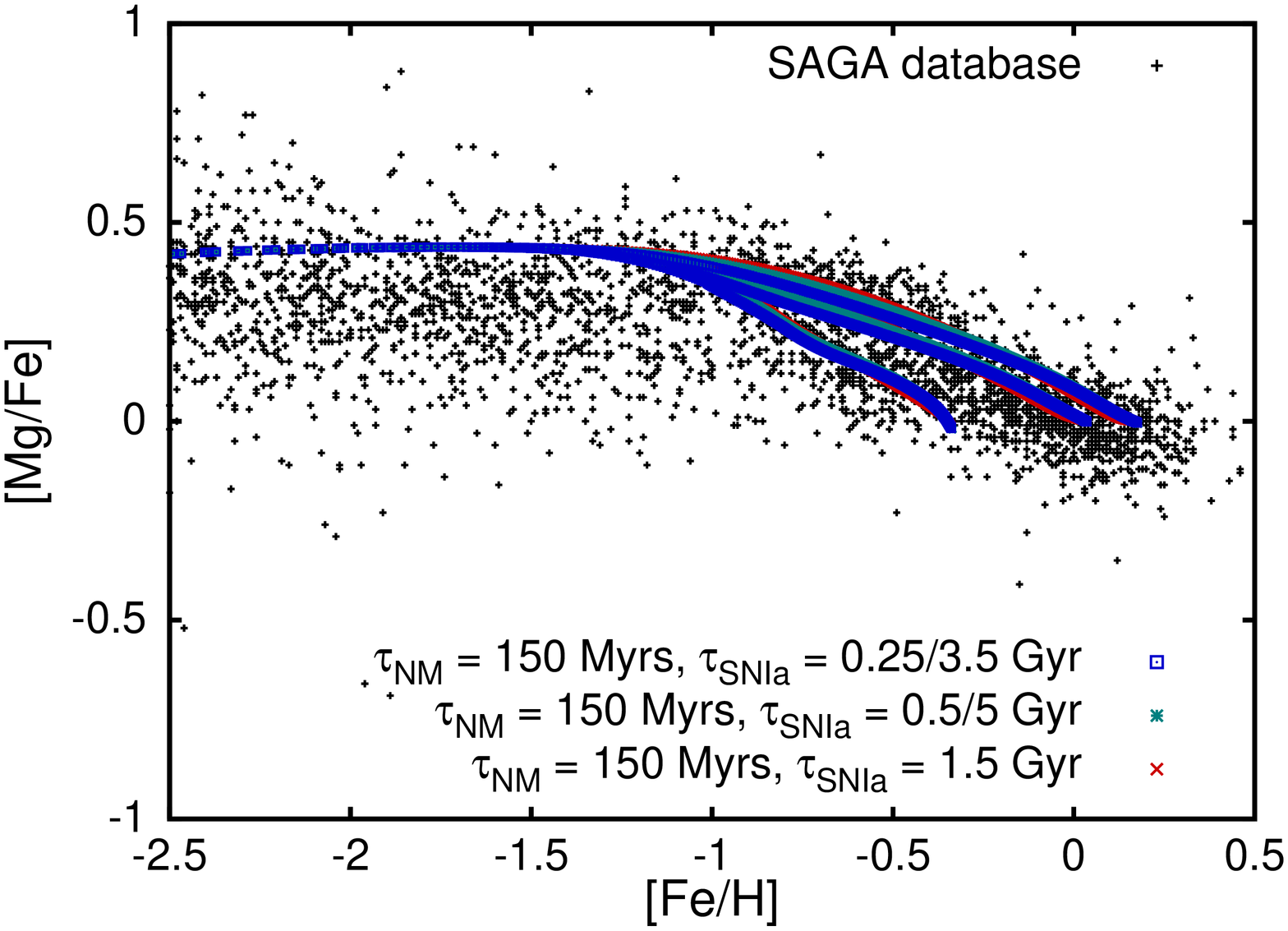,angle=-90,width=0.49\hsize}
\caption{Testing different SNIa delay time distributions. The top row shows $\eufe$/$\rfe$, the bottom row displays $\mgfe$, while the left-hand column contrasts a 1-phase chemical evolution model with a model including a hot phase with $\fcNM=\fcccSN=0.25$ in the right-hand column. The different colours encode trajectories for different delay time distributions for the SNeIa: A single exponential law with $\tau_\snIa = 1.5 \Gyr$, and two double exponential laws with equal weight for both components and $\tau_\snIa = 0.25,3.5 \Gyr$ (fast) and $\tau_\snIa = 0.5,5 \Gyr$ (slow).}\label{fig:modeldtd}
\end{figure*}

\begin{figure}
\epsfig{file=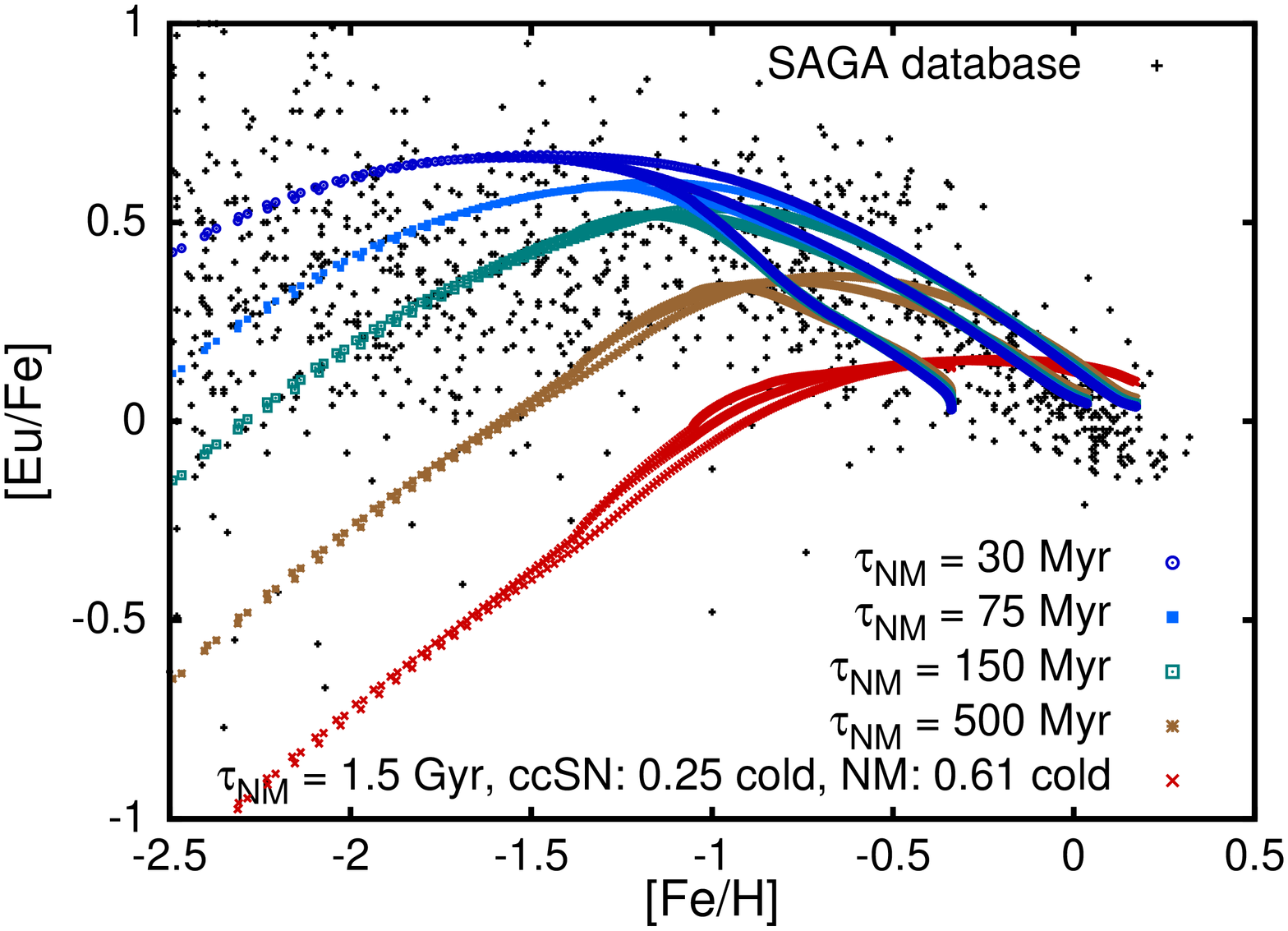,angle=-90,width=\hsize}
\caption{Testing different hot neutron star merger characteristic delay timescales $\tau_\NM$ for the standard model with $\fcccSN=0.25$ and $\fcNM = 0.61$}\label{fig:modeltNM}
\end{figure}

\section*{Appendix}

\subsection*{A1 Analytical model}\label{sec:FullAnalytical}

In this section we develop a more sophisticated analytical model for Eu vs. Mg enhancement. As an approximate guide to the behaviour of $\eumg$ abundance ratios with a 2-phase ISM, we can adapt equations (34) and (37) of Weinberg et al. (2017, hereafter WAF), which describe the abundance evolution for an element with instantaneous enrichment and an element whose production follows an exponential delay time distribution with a minimum time delay $\tmin$.

These equations assume a constant SFR, a constant star formation efficiency timescale $\tau_* \equiv M_g/\mdotstar$, and outflow with a constant mass-loading factor $\eta \equiv \mdotout/\mdotstar$.
With some notational changes the enrichment in the instantaneous and delayed cases can be written 
\begin{equation}
  Z_{\rm inst}(t) = {y_{\rm inst} \over 1+\eta-r}\left[1-e^{-t/\taudep}\right]
  \label{eqn:zinst}
\end{equation}
and
\begin{equation}
  Z_{\rm del}(t) = {y_{\rm del} \over 1+\eta-r}\left[1-e^{-\frac{\Delta t}{\taudep}}
	- {\taudelay \over \taudelay-\taudep}(e^{-\frac{\Delta t}{\taudelay}} -
  e^{-\frac{\Delta t}{\taudep}})\right]~,
  \label{eqn:zdelay}
\end{equation}
where $r \approx 0.4$ is the recycling fraction, $\Delta t = t-\tmin$, and $\taudep = \tau_*/(1+\eta-r)$ is the timescale on which the gas supply would be depleted (in the absence of accretion) by the combination  of star formation and outflow. For present purposes, we use equation~(\ref{eqn:zinst}) multiplied by $\fcccSN$ to represent the ccSN Mg enrichment direct to the cold phase, and we use equation~(\ref{eqn:zdelay}) multiplied by $(1-\fcccSN)$ to represent ccSN Mg enrichment that passes through the hot phase, with $\taudelay = \taucool$.
We use equation~(\ref{eqn:zdelay}) multiplied by $\fcNM$ to represent NM Eu enrichment direct to the cold phase, with $\taudelay=\tauNM$ and a minimum delay time $\tminNM$, and we use equation~(\ref{eqn:zdelay}) multiplied by $(1-\fcNM)$ and with timescale $\taudelay=\tauNM+\taucool$ to represent NM Eu enrichment  that passes through the hot phase.
In similar fashion, we can compute the ccSN and SNIa contributions to iron enrichment.

Although the assumptions and parameters of this approximate analytic model differ from those of our numerical model, we find that it reproduces the qualitative behaviour of the numerical $\eumg$ and $\eufe$ results for different choices of $\tauNM$ and $\fcNM$.

For typical yield datasets \citep[][or the yields used here]{Andrews17}, an outflow mass-loading $\eta \approx 2.5$ is needed to obtain approximately solar abundances at late times. This factor would be moderately modified in the presence of radial flows. Figure~\ref{fig:analytic}a shows the $\mgfe$ evolution (dotted curve) and $\eufe$ evolution (solid curves) for a model with $\eta=2.5$, $\tau_*=1\Gyr$, $\fcNM=\fcccSN=0.25$, and a range of choices for $\tauNM$.  We have adjusted the ccSN iron yield to give a plateau at $\mgfe_{\rm pl} \approx 0.45$ similar to the numerically solved chemical evolution in Section~\ref{sec:specmodel}, and we have used the same SNIa parameters ($\fcSNIa =0.01$, $\tau_\SNIa=1.5\Gyr$, minimum time delay $0.15\Gyr$), cooling time $\tau_c=1\Gyr$, and minimum NM delay ($0.015\Gyr$) as the numerical models. Comparison to Figure~\ref{fig:modelnohot} shows similar trends with $\tauNM$, similar peak values of $\eufe$, and similar evolution for $\feh \ga -1.5$.
At times $t \gg \tauNM$, ccSN and NM enrichment proceed at equal rates, so all $\eufe$ curves asymptote to match the $\mgfe$ curve, the shape of which is driven by SNIa iron enrichment.

At early times and low $\feh$, the analytic predictions of $\eufe$ in Figure~\ref{fig:analytic}a are significantly lower than those in Figure~\ref{fig:modelnohot}, primarily owing to the different treatment of outflows. The model presented in the main text assumes a fixed loss of stellar yields ($60 \%$) vs. the mass loss ($\eta$) formalism applied in this section, which drives an outflow at the ISM metallicity. In the analytic model, less metals are lost at early times, when the ISM metallicity is low. The ccSN enrichment of the $\eta = 2.5$ model therefore proceeds faster at early times, moving the initial rise in $\eumg$ and $\eufe$ to the right.  We replicate the difference between the models by multiplying all yields by a factor of 0.4 and setting $\eta=0$ in Figure~\ref{fig:analytic}b, which as expected shifts the early rise in $\eufe$ back to the left, approaching Figure~\ref{fig:modelnohot}, though if we changed to a single-phase model ($\fcccSN=\fcNM=\fcSNIa=1$) then the early rise would still be slower than in Figure~\ref{fig:modelnohot}. The remaining differences may be explained by the instantaneous return of locked up metals in the analytic approach of this Section vs. the correctly delayed return allowed by the numerical solution. Radial gas flows in the model of Section~\ref{sec:specmodel} depress the local metallicity, so that its net mass loss rate is smaller. Most important for our purposes, the maximum values of $\eufe$ are nearly identical for both outflow implementations.  

Figures~\ref{fig:analytic}c and~\ref{fig:analytic}d show varying values of $\fcNM$ with $\tauNM$ fixed to $0.15\Gyr$, for the same $\eta$ and yield combinations in panels (a) and (b), respectively. Here we adopt $\fcccSN=0.25$, as in the left column of Figure~\ref{fig:modelspecvar}. The behaviour of the analytic models resembles that of the  numerical models, though the values of $\eufe$ are again lower at low $\feh$, and the maximum values of $\eufe$ are slightly lower (by $\sim 0.05$ dex). The final panels adopt  $\fcNM=0.50$ and $\tauNM=0.15\Gyr$ and vary the outflow mass-loading $\eta$ or the hot-phase cooling time $\tau_c$. Changing $\eta$ alters the final $\feh$ but not the maximum of $\eufe$.  Shortening $\tau_c$ to $0.1\Gyr$ allows hot-phase enrichment to occur much more quickly, so the model has already evolved to $\feh \approx -0.5$ before $\eufe$ reaches its maximum.  However, models with $\tau_c=0.5$ or $2.0\Gyr$ are only slightly different from our standard case of $1.0\Gyr$. The results in Fig.~\ref{fig:analytic}f qualitatively resemble those in Fig.~\ref{fig:modeltcool}.

We can use the analytic model to gain some insights into behaviour in different regimes. For a star formation efficiency timescale $\tau_*=1\Gyr$ and $\eta=2.5$, the depletion timescale is $\tau_d \approx 0.3\Gyr$.  For $\tau_c \approx 1\Gyr$, we expect $\tau_c > \tau_d > \tauNM$.  If we make the approximation $\taucool \gg \taudep \gg \tauNM$, implying $\taucool/(\taucool-\taudep) \approx 1$ and $\tauNM/(\tauNM-\taudep) \approx 0$, we get
\begin{equation}
  {Z_{\rm Eu}(t) \over Z_{\rm Mg}(t)} \approx 
  \left({y_{\rm Eu} \over y_{\rm Mg}}\right)
  {\fcNM F(t,\taudep) +
    (1-\fcNM)F(t,\taucool)
	\over
    \fcccSN F(t,\taudep)+
	(1-\fcccSN) F(t,\taucool)}
   \label{eqn:zratio} $,$
\end{equation}
where 
\begin{equation}
F(t,\tau) = 1 - e^{-t/\tau} $.$
\end{equation}
When $t \ll \taucool$ so that the first terms dominate, the abundance ratio is equal to the yield ratio multiplied by $\fcNM/\fcccSN$.  When $t \gg \taucool$, the abundance ratio is simply equal to the yield ratio, which should be similar to the solar Eu/Mg ratio if the model is to reproduce observations.

The approximation that leads to equation~(\ref{eqn:zratio}) breaks down when $t \ll \tauNM$.  In this regime, one can instead ignore the hot phase return and use a Taylor expansion of
equations~(\ref{eqn:zinst}) and~(\ref{eqn:zdelay}) to find 
\begin{equation}
  {Z_{\rm Eu}(t) \over Z_{\rm Mg}(t)} \approx
  {1 \over 2} \left({y_{\rm Eu} \over y_{\rm Mg}}\right)
	\left({\fcNM \over \fcccSN}\right)
	\left({t \over \tauNM}\right)
	\left({1-{\left(\frac{\tminNM}{t}\right)}^2}\right)
	\label{eqn:early}
\end{equation}
(for $t > \tminNM$).
Thus, the Eu/Mg ratio grows approximately linearly at early times, leveling off as $t$ approaches $\tauNM$. By this time, the Mg abundance has reached a fraction
$\fcccSN (1-e^{-\tauNM/\taudep}) \approx \fcccSN(\tauNM/\taudep)$ of its final equilibrium value, assuming that hot phase return is still negligible. From these analytic arguments, we conclude that $\fcNM > \fcccSN$ is required to produce [Eu/Mg]$>0$ at any time and that the condition for reaching [Eu/Mg]$>0$ by the time [Mg/H]$\approx -1$
is approximately $\fcccSN(\tauNM/\taudep) < 0.1$. Both conditions are physically plausible. Reaching [Eu/Mg]$>0$ while [Mg/Fe] is significantly super-solar requires $\tauNM < \tau_{\rm Ia}$, which is again physically plausible.

\subsection*{A2 Further model tests}\label{sec:SNIaDTD}

In Fig.~\ref{fig:modeldtd} we present models with a 1-phase ISM (left-hand column) and a 2-phase ISM (right-hand side) for three different SNIa DTDs, as discussed in \cite{Weinberg17}. These double-exponential DTDs are designed to more closely match the frequently claimed $t^{-1}$ time-dependence, while they preserve the straight-forward normalisation and integrability of exponential functions.

While there might be some targeted applications where this matters, the short message for our work is: it does not significantly impact our results.

Finally, Fig.~\ref{fig:modeltNM} shows different characteristic delay timescales $\tau_\NM$ for models with our standard cooling timescale of $\tau_{\rm cool} = 1 \Gyr$ and more yields from neutron star mergers directly entering the cold gas phase ($\fcNM = 0.61$ vs. $\fcccSN = 0.25$. The behaviour is as expected, with the models with shorter $\tau_\NM$ showing faster $\eu$ enrichment and thus reaching a larger peak $\eufe$ value, before SNeIa and cooling of less r-process rich material from the hot phase set in.

\label{lastpage}
\end{document}